\documentclass[12pt]{article}

\usepackage{cite,graphicx,amsmath,amssymb,bbm}
\usepackage{floatflt,epsf,pifont,amsfonts,mathrsfs}

\newcommand{\be}{\begin{equation}}
\newcommand{\ee}{\end{equation}}
\newcommand{\bea}{\begin{eqnarray}}
\newcommand{\eea}{\end{eqnarray}}

\newcommand{\dec}{{\rm \text{\tiny dec}}}
\newcommand{\IRc}{{\rm \text{\tiny IR,cr}}}
\newcommand{\IR}{{\rm \text{\tiny IR}}}
\newcommand{\UV}{{\rm \text{\tiny UV}}}
\newcommand{\RH}{{\rm \text{\tiny RH}}}
\newcommand{\pt}{{\rm \text{\tiny pt}}}

\newcommand{\eff}{{\rm \text{\tiny eff}}}

\begin{document}

\thispagestyle{empty}
\vspace*{.2cm}
\noindent
HD-THEP-08-07 \hfill 25 January 2008

\vspace*{1.5cm}

\begin{center}

{\Large\bf Sequestered Dark Matter}\\[2.5cm]

{\large B.~v.~Harling and A.~Hebecker}\\[.5cm] {\it
Institut f\"ur Theoretische Physik, Universit\"at Heidelberg,
Philosophenweg 16 und 19, D-69120 Heidelberg, Germany} \\[.5cm]
{\small\tt (\,harling} {\small and} {\small\tt
a.hebecker@thphys.uni-heidelberg.de)} \\[2.0cm]

{\bf Abstract}
\end{center} 
\noindent 
We show that hidden-sector dark matter is a generic feature of the type
IIB string theory landscape and that its lifetime may allow for a
discovery through the observation of very energetic $\gamma$-rays produced in 
the decay. Throats or, equivalently, conformally sequestered hidden sectors 
are common in flux compactifications and the energy deposited in 
these sectors can be calculated if the reheating temperature of the standard 
model sector is known. Assuming that throats with various warp factors are 
available in the compact manifold, we determine which throats maximize 
the late-time abundance of sequestered dark matter. For such throats, this 
abundance agrees with cosmological data if the standard model reheating 
temperature was $10^{10} - 10^{11}$~GeV. In two distinct scenarios, the 
mass of dark matter particles, i.e. the IR scale of the throat, is either 
around $10^5$~GeV or around $10^{10}$~GeV. The lifetime and the decay
channels of our dark matter candidates depend crucially on the fact that the 
Klebanov-Strassler throat is supersymmetric. Furthermore, the details of 
supersymmetry breaking both in the throat and in the visible sector play an 
essential role. We identify a number of scenarios where this type of dark 
matter can be discovered via $\gamma$-ray observations.

\newpage
\section{Introduction}
Dark matter is frequently assumed to consist of massive weakly interacting 
particles which are stable (or have a very long lifetime) because their 
decay is forbidden by some (approximate) symmetry. However, it is also 
well-known that dark matter may originate in a hidden sector which is coupled 
to the standard model only via higher-dimension operators, ensuring 
that dark matter does not decay (see e.g.~\cite{Kolb:1985bf,smallnumbers}). 

We demonstrate that the latter scenario is realized under fairly general 
assumptions in the type IIB string theory landscape. The main starting point 
is the well-known fact that type IIB flux compactifications generically 
contain many strongly warped regions or throats~\cite{Denef:2004ze}. This 
statement can be made quantitative~\cite{Hebecker:2006bn} under the 
assumption that the fine-tuning of the cosmological constant requires 
manifolds with many 3-cycles and that, in many cases, such 3-cycles produce 
Klebanov-Strassler throats~\cite{KS} if stabilized at small volume. We base
our analysis on the fact that, after inflation ends, the standard model 
sector is heated to a certain temperature. Moreover, we assume that the 
throats have received no energy from the reheating process.\footnote{
Alternatively, 
if this assumption is not fulfilled, i.e.~if the throats are heated directly 
by the reheating mechanism, a lower reheating temperature would be sufficient 
for our scenario.
} 
Even under such minimal assumptions, a certain amount of energy is deposited 
in the various throats by energy transfer from the heated standard model 
sector. There exist certain optimal throat lengths (i.e. optimal warp
factors) 
for which a given throat contains a maximal amount of cold dark matter in the 
late universe. Requiring furthermore that such an `optimal' throat is 
responsible for the dark matter observed today, we determine the reheating 
temperature to be between $ 10^{10}$~GeV and $10^{11}$~GeV. The IR scale of 
the relevant throat (i.e. the mass of dark matter particles) is either around 
$10^5$~GeV or around $10^{10}$~GeV. Moreover, we observe that the decay rate
of
throat dark matter to the standard model sector is not negligible. It may, for
a certain range of parameters, lead to the discovery of this variant of dark
matter via the observation of the diffuse $\gamma$-ray background or the
spectrum of very energetic $\gamma$-rays. 

The fact that dark matter can come from a hidden (or more precisely 
conformally sequestered) sector realized by a Klebanov-Strassler throat has
already been emphasized in~\cite{CT}. This paper focuses on scenarios where 
a Klebanov-Strassler throat is heated by the annihilation of a brane with an
antibrane at the end of inflation. Subsequently, energy is transferred from
this throat to other throats which may be present in the compact manifold.
Throat-localized Kaluza-Klein (KK) modes, which are produced in this way, can
be the observed dark matter if they are sufficiently long-lived and if certain
parameters which determine their relic density are tuned. More
precisely, three types of dark matter candidates are discussed in \cite{CT}:
Kaluza-Klein modes which are localized in the throat where inflation took
place and those localized in the throat where the standard model lives have to
carry an (approximately) conserved angular momentum in the throat in order to
be sufficiently stable. Kaluza-Klein modes which are localized in other
throats may, by contrast, be long-lived also without such an angular 
momentum.\footnote{
In 
addition, \cite{CT} also discusses particles on D-branes in these throats as 
a dark matter candidate.}

We approach the possibility of throat dark matter from a different and, to a 
certain extent, more general perspective: We do not assume reheating to be 
due to brane-antibrane annihilation in a warped region. Instead, we only rely 
on the fact that the standard model (which is realized in the unwarped part
of the manifold) has a certain reheating temperature after 
inflation ends. In our approach, throats are not a `model building feature'
introduced to realize inflation, uplifting, etc. Instead, we view the presence 
of (a potentially large number of) throats as a prediction of the type IIB 
string theory landscape. Accordingly, we consider scenarios in which various 
throats are present and perform a quantitative analysis of the energy density 
in these throats in the late universe. We attempt to avoid any other specific
assumptions, relying only on the known fact that the standard model has a 
certain initial temperature in early cosmology. The resulting dark matter 
can then be viewed as a generic prediction of the type IIB landscape. It 
turns out that, given a reasonably large number of throats, the reheating 
temperature is the only parameter which has to be tuned to account for the 
observed dark matter density. We show quantitatively, using results 
from~\cite{Hebecker:2006bn}, that sufficiently many throats are available in 
an important fraction of vacua of the type IIB landscape.

We focus in particular on the phenomenological importance of fermionic KK 
modes in the throat. To the best of our knowledge, this point has so far not 
received sufficient attention in the literature. The Klebanov-Strassler 
throat has $\mathcal{N}=1$ supersymmetry. Generically, this supersymmetry 
is weakly broken by the compact space to which the throat is attached and 
in which supersymmetry has to be broken for phenomenological reasons. 
Therefore, light fermionic KK modes are guaranteed to be present in the 
spectrum. They turn out to play a central role in the phenomenology of 
sequestered dark matter. Furthermore, we use our new values for the energy 
transfer rates and the decay rates between throats~\cite{hotthroats}. For 
temperatures and Kaluza-Klein masses smaller than the compactification scale, 
our rates differ from the values used in~\cite{CT}. We restrict our
ana\-lysis 
to this case, which can also be characterized as the case of `overlapping 
potentials' in the quantum mechanical language of tunneling~\cite{hotthroats}. 
Our reasons for focusing on this parameter range are as follows: On the one 
hand, a large compactification scale is required to make the KK modes
sufficiently stable against decay to the standard model or other
throats.\footnote{
An exception is the decay rate of fermionic KK modes to
the standard model sector which can be made small even for small
compactification scales (cf.~Sect.~\ref{DecaytoSM}).} 
On the other hand,
temperatures above the compactification scale would imply that also the
unwarped part of the manifold is heated up. The resulting gas of KK modes in
the compact space may then destabilize the volume modulus, making the
analysis of the early cosmological evolution much more difficult.

Our paper is organized as follows: Sect.~\ref{thermalproduction} describes the
thermal production of sequestered dark matter. Using the energy transfer rates
from~\cite{hotthroats}, we determine the energy density
deposited in a throat by the heated standard model sector. We show that the
KK modes which are produced in that way thermalize for a
certain range of parameters. Whether this happens or not influences
the further time evolution of the energy density significantly. Taking this
into account, we calculate the late-time abundance of KK modes as a function
of the
reheating temperature and the IR scale of the throat. In
Sect.~\ref{candidates}, we discuss various decay channels of the KK modes. We
show that they decay very quickly to a scalar state and its
fermionic superpartner. These lightest KK modes can then decay to the
standard model
sector. We determine the corresponding decay rates which turn out to be
different for the scalar and the fermion. Cosmological scenarios are discussed
in Sect.~\ref{scenarios}. First, we analyse setups with a single throat. We
find that a moderately long throat gives a promising dark matter candidate
which may allow for a discovery by upcoming $\gamma$-ray experiments.
Then, we consider scenarios with a large number of throats, using results on
the distribution of multi-throat configurations from~\cite{Hebecker:2006bn}.
We find that a throat of the required length is in many cases present. Some
issues concerning supersymmetry breaking in the throat sector are
discussed in Sect.~\ref{extra}. Finally, in Sect.~\ref{summary}, we give a
summary of our results.

\section{Thermal production}
\label{thermalproduction}

\subsection{Energy transfer}\label{et}
As outlined in the Introduction, we assume the throats to have received 
no energy from the reheating process, whereas the standard model is 
heated to a temperature $T_{\RH}$ initially. Subsequently, energy will 
be transferred from the standard model to the throats. Note that this 
process is similar to the energy transfer from the hot brane to the bulk in 
Randall-Sundrum II models~\cite{Hebecker:2001nv} (see also~\cite{ 
Gubser:1999vj}). The AdS$_5$ bulk plays the role of the throat, which we 
however assume to be of finite length, with the Klebanov-Strassler region 
corresponding to the IR brane. 

In~\cite{hotthroats}, we have estimated the energy transfer rate between a 
heated Klebanov-Strassler (KS) throat at temperature $T$ and another throat in
a type IIB flux compactification. Our estimate is based on a calculation 
which can be performed in a simpler setting: two AdS$_5 \times$S$^5$ throats 
embedded in a 6d torus. For the purpose of this analysis, we have replaced 
the throats by equivalent stacks of D3-branes. In this language, the 
aforementioned energy transfer is that from a heated gauge theory living on 
one brane stack via supergravity fields in the embedding torus to the gauge 
theory living on the other brane stack. The coupling of supergravity in the 
torus to the gauge theories on the two brane stacks follows from the DBI 
action. Performing a KK expansion of the supergravity
fields, the calculation of the energy transfer rate becomes a simple exercise
in quantum field theory. One finds
\be
\label{tr1}
\dot{\rho} \sim \frac{N_1^2 N_2^2}{M_{10}^{16} A^8} \, T^{13} +
\frac{N_1^2 N_2^2}{M_{10}^{16} L^{12}} \, T^9,
\ee
where $M_{10}$ is the 10d Planck scale, $A$ is the distance between the two 
brane stacks/throats and $L$ is the size of the embedding torus. Moreover, 
$N_1$ and $N_2$ are the numbers of branes of the two stacks. The world-volume 
theories then have $\sim N_1^2$ and $\sim N_2^2$ degrees of freedom, 
respectively.

This energy transfer rate is also applicable to our setup. To this end, 
instead of a heated throat, we consider some D3- and/or D7-brane-realization 
of the standard model (with $g \sim 100$ degrees of freedom). We assume
these D-branes to live in the unwarped part of the manifold (and not in a
throat). To account for the standard model, we simply have to set $N_1^2 = g$
in Eq.~\eqref{tr1}. Moreover, instead of an AdS$_5 \times$S$^5$ throat we are
interested in energy transfer to a KS throat. The equivalent description is
that of a stack of D3- and fractional D3-branes at a conifold singularity. For
the absorption process, the relevant part of the geometry is the UV end of
the throat, which is well approximated by AdS$_5 \times$T$\smash{^{1,1}}$.
This in turn is equivalent to a large number of D3-branes at a 
conifold singularity, which we denote by $N_\UV$ since it corresponds to 
the number of 5-form flux at the UV end of the KS throat. Therefore, 
$N_2 = N_\UV$ in Eq.~\eqref{tr1}. We assume that the temperature of the 
standard model is smaller than the compactification scale, i.e.~$\smash{T < 
L^{-1}}$. Moreover, we consider the generic situation that the distance 
between the two brane stacks is of the same order of magnitude as the size 
of the embedding manifold, i.e.~$A \sim L$. In this case, the second term in 
Eq.~\eqref{tr1} dominates.\footnote{
All 
subsequent results are easily extended to the case of nearby throats by 
simply using the first instead of the second term in Eq.~\eqref{tr1} and 
performing an analogous modification of the dark matter decay rates.
}
As explained in~\cite{hotthroats}, this term is due to the effect of the zero 
mode in the KK expansion of supergravity fields and is therefore completely 
insensitive to the unknown details of the Calabi-Yau geometry. In~\cite{
hotthroats}, we focused on the dilaton, which is one of the supergravity 
fields mediating the energy transfer. However, using the coupling of the 10d
graviton to the energy-momentum tensors on the two brane stacks, 
Eq.~\eqref{tr1} can also be derived on the basis of the graviton as the 
mediating field. Since the zero mode of the dilaton acquires a mass in flux 
compactifications \`a la GKP~\cite{Giddings:2001yu}, its effect is suppressed 
in this case. We therefore focus on the graviton zero mode as the field 
responsible for the second term in Eq.~\eqref{tr1}. 

Using the relation $\smash{M_4 \sim M_{10}^4 L^3}$ for the 4d Planck scale,
the energy transfer rate is given by
\be
\label{etr}
\dot{\rho} \, \sim \, g \, N^2_{\UV} \frac{T^9}{M_4^4}.
\ee
This rate is easily understood as being due to a gravitational strength 
coupling between a sector with $g$ degrees of freedom (the standard model) 
and a sector with $N_\UV^2$ degrees of freedom (the KS throat). 

The finite length of the KS throat implies the existence of an IR scale
$m_\IR$, the mass of the lowest-lying KK mode in the throat. In the dual
picture, this corresponds to the fact that the fractional D3-branes break 
conformal invariance and induce a confinement or IR scale $m_\IR$ in the 
gauge theory. Thus, KK modes in the throat (i.e. glueballs of the dual gauge
theory) can only be created if $T_\RH>m_\IR$. We then expect these glueballs
to have masses up to $m \sim T_\RH$. 

The glueballs may decay back to the standard model. Jumping somewhat 
ahead, we note that spin-2 glueballs have the highest decay rate 
(see Sect.~\ref{DecaytoSM} for details):
\be
\label{KK2SM}
\Gamma(m) \, \sim \, g \, N^2_{\UV} \frac{m^4 m_\IR}{M_4^4}.
\ee
On the other hand, glueballs can also decay to lighter glueballs within the 
same throat. We will have to discuss this process in some detail below. At 
the moment, it is sufficient to establish that the decay to lighter glueballs
wins over the possible decay back to the standard model. For this purpose, we
recall that we are dealing with a strongly coupled system with a dense 
spectrum. Thus, the initially created gauge theory state of mass $m$ will 
have a lifetime $\sim 1/m$. In the most conservative scenario, it will decay 
to 2 states of mass $m/2$. These states will in turn decay to states of mass 
$m/2^2$ after a time-interval $\sim 2/m$, and so on. Summing up the 
probabilities for the decay back to the standard model at each step of this 
cascade, we arrive at a total probability 
\be 
w\sim\sum_{n=0} \Gamma(m/2^n)\cdot\frac{2^n}{m}\,.\label{w}
\ee
This sum is of the same order of magnitude as the first term and hence very 
small in all cases of interest. Clearly, we could equally well have assumed 
that each glueball decays to $k_1$ lighter states with mass $m/k_2$, arriving 
at the same conclusion for any ${\cal O}(1)$ numbers $k_1$ and $k_2$. Thus, 
the relaxation to lighter states within the same throat always wins over the 
decay back to the standard model or to other throats. 

The energy transfer rate Eq.~\eqref{etr} is strongly temperature dependent, 
$\dot{\rho} \propto T^9$. Therefore, energy transfer is effectively finished 
soon after reheating and the corresponding time scale is $\smash{|T/\dot{T}|}$ 
at $T=T_\RH$. The total energy density after reheating is dominated by the 
relativistic gas in the standard model sector with $\smash{\rho = g 
\frac{\pi^2}{30} T^4}$. We find
\be
\label{rts}
|T/\dot{T}| \, = \, H^{-1} \, \sim \,  \frac{M_4}{ g^{1/2}T^2},  
\ee
where $H$ is the Hubble rate. Using Eqs.~\eqref{etr} and \eqref{rts} at $T 
\sim T_\RH$, the energy density deposited in a throat directly after 
reheating is 
\be
\label{energydensity}
\rho \, \sim \, \dot{\rho} \, |T/\dot{T}| \, \sim \, g^{1/2} 
N^2_\UV \frac{T^7_\RH}{M_4^3}.
\ee

Before closing this section, we note that the energy transfer processes 
we consider compete with the unavoidable energy deposition in the throat 
sectors occurring during inflation. This can be understood by noting that 
de Sitter space has a temperature $T_{\rm dS} \sim 1/R_{\rm dS}\sim 
M_{\rm Inf}^2/M_4$. We assume that inflation lasts long enough for 
the throats to be thermalized with this temperature. Furthermore, 
parameterizing the efficiency of reheating by an efficiency factor
$\epsilon\le 1$, we have $gT_{\rm RH}^4\sim \epsilon M_{\rm Inf}^4$. Thus, 
all throats have a temperature $T_{\rm dS} \sim \sqrt{g/\epsilon}\,T_{\rm 
RH}^2/M_4$ at the time of reheating. Jumping ahead, we note that for 
typical long throats (where this effect is most relevant), we find 
initial throat temperatures $\sim 10^6$~GeV and $T_{\rm RH}\sim 10^{11}$~GeV. 
For such throats, `de-Sitter heating' in fact wins over the heating process 
analysed in this section if $\epsilon<1$, allowing in principle for even 
more throat dark matter than we find in our conservative analysis.

\subsection{Time evolution of the energy density}
\label{te}
The gauge theory states created by energy transfer from the standard model 
sector decay into a certain number of the lightest glueballs of mass $m_\IR$ 
with a certain distribution of kinetic energies. Two extremal cases are 
possible. In one case, the initial gauge theory state decays into an 
$\mathcal{O}(1)$ number of the lightest glueballs which accordingly have 
kinetic energies of the order of $T_\RH$. In the other case, the initial 
gauge theory state settles into a large number of the lightest glueballs with 
kinetic energies of the order of their mass $m_\IR$. In between these two 
extremal cases, a continuous distribution of kinetic energies is a priori 
possible. The knowledge of this distribution is important since it determines 
whether the glueballs can reach thermal equilibrium after reheating and 
whether (or for how long) the energy density scales like radiation or like 
matter with the expansion of the universe. This in turn determines how large 
the contribution of the glueballs to the total energy density is at our 
epoch. Since we are at present unable to determine this distribution of 
kinetic energies, we discuss the two extremal cases separately. This allows 
us to estimate the possible range of contributions of the glueballs to 
the current energy density of the universe as a function of the parameters 
$m_\IR$, $N_\UV$ and $T_\RH$.

We begin by collecting some results from the literature on the thermodynamics 
of the gauge theories that are relevant for our discussion. The gauge theory 
dual to a KS throat has a logarithmically varying number of degrees of 
freedom, corresponding to the logarithmic deviation of the KS geometry from 
AdS$_5$. In the deconfined phase, the effective number of colours $N_\eff$ of 
the gauge theory depends on the temperature $\smash{\tilde{T}}$ of the plasma 
as
\be
\label{N_eff}
N_\eff \, \sim \, N_\IR \, \ln \left( \frac{\tilde{T}}{m_\IR} \right).
\ee
Here, $N_\IR $ is the number of 5-form flux at the bottom of the 
throat. The deconfined phase of the gauge theory is dual to a throat with a
black hole horizon which replaces the IR end. The highest meaningful value in
Eq.~\eqref{N_eff} is $N_\eff \sim N_\UV$. This corresponds to a temperature
where the black hole horizon reaches the UV end of the throat. The energy
density as a function of the plasma temperature $\smash{\tilde{T}}$ is 
\be
\rho \, \sim \, N_\eff^2 \, \tilde{T}^4.
\ee
Since the logarithmic variation of $N_\eff$ with $\smash{\tilde{T}}$ is small
compared to the variation of the $\smash{\tilde{T}^4}$-term, we will neglect 
it in the following. The deconfined phase can then be described by an 
approximate conformal field theory and the energy density correspondingly 
scales like radiation with $\smash{a^{-4}}$, where $a$ is the scale factor of 
the universe. When the energy
density drops to $\rho \sim N^2_\IR m^4_\IR$, a confinement phase transition
begins which lasts until the energy density has reached $\rho \sim g_s N_\IR
m_\IR^4$ (see e.g.~Appendix A in~\cite{Aharony:2005bm}; the phase transition
in the dual gravity picture was studied in~\cite{BHinThroat1} using a 5d 
picture and in~\cite{BHinThroat2} using the full KS geometry). In the 
transition region for $\rho$, space is divided into separate regions in 
either the confined
phase with $\rho < g_s N_\IR m_\IR^4$ or the (still) deconfined phase with 
$\rho > N^2_\IR m^4_\IR$. At even lower energy densities $\rho < m_\IR^4$ 
(assuming $g_s N_\IR >1$), a description in terms of a nonrelativistic 
glueball gas is applicable and the energy density correspondingly scales with 
$\smash{a^{-3}}$. We do not know the scaling of $\rho$ with $a$ in the 
transition region $\smash{N^2_\IR m^4_\IR  > \rho > m_\IR^4}$ though, since 
the equation of state during the phase transition is unknown. Since we expect 
the scaling to be in between the two extremes $\smash{\rho \propto a^{-3}}$ 
and $\smash{\rho \propto a^{-4}}$ in this region, we will take 
$\smash{\rho \propto a^{-4}}$ for $\smash{\rho > N_\IR m_\IR^4}$ and 
$\smash{\rho \propto a^{-3}}$ for $\smash{\rho < N_\IR m_\IR^4}$ for 
simplicity.\footnote{
Using 
the intermediate value $\smash{\rho \sim N_\IR m_\IR^4}$ for the distinction 
between the two behaviours, the error in $\rho$ is a factor of $\smash{(N_\IR 
m_\IR^4 / N^2_\IR m_\IR^4)^{1/4} = N_\IR^{-1/4}}$ if $\smash{\rho \propto 
a^{-3}}$ in the entire transition region or a factor of $\smash{(N_\IR 
m_\IR^4 / m_\IR^4 )^{1/3} = N_\IR^{1/3}}$ if $\smash{\rho \propto a^{-4}}$ in 
the entire transition region. In both cases, this factor is typically 
$\mathcal{O}(1)$.}

Let us first consider the case that the energy density deposited in a throat, 
Eq.~\eqref{energydensity}, is larger than $\rho \sim N_\IR m_\IR^4$. If the 
gauge theory state created at reheating decays into a large number of 
glueballs with mass and kinetic energy of the order of $m_\IR$, the gauge 
theory thermalizes. To see this in more detail, we view each initially 
created glueball as a localized excitation of a strongly 
coupled system with energy $\sim T_\RH$. The localization assumption can be 
justified by recalling that, from the D-brane perspective, the mediating 
bulk supergravity fields couple to local gauge theory operators like 
$F_{\mu\nu}F^{\mu\nu}$. We model the further evolution of this state as a 
ball of gauge theory plasma expanding with the velocity of light.\footnote{
Note 
that this physical picture is equivalent to the picture of a cascade decay 
used in the derivation of Eq.~(\ref{w}) if we assume that a glueball with mass 
$m/2^n$ fills out a volume $(2^n/m)^3$. 
}
The number density of these balls is
\be
\label{nd}
n \, \sim \frac{\rho}{T_\RH} \, \sim \, g^{1/2} N^2_\UV  \frac{T_\RH^6}{M_4^3},
\ee
where we have used Eq.~\eqref{energydensity}. It follows that the balls
fill out the whole space after a time
\be
t \, \sim \, n^{-1/3} \, \sim \, \frac{M_4}{g^{1/6}N^{2/3}_\UV \, T_\RH^2}\,.
\ee 
Comparing this with the Hubble time Eq.~\eqref{rts} at $T=T_\RH$, we see that
the gauge theory plasma fills out the whole space before the Hubble expansion 
becomes relevant if $N_\UV^2 \gtrsim g$, which holds for all relevant throats.

The question of thermalization is more subtle if the gauge theory state 
produced at reheating decays into an $\mathcal{O}(1)$ number of the lightest 
glueballs with kinetic energies of the order of $T_\RH$. We defer the 
corresponding discussion to Sect.~\ref{subtleties}. It turns out that in the 
cases of interest the glueballs again thermalize if the initial energy density 
is larger than $\rho \sim N_\IR m_\IR^4$. Furthermore, the initial decay 
vertex is strong enough to ensure that the potential energy $\sim m$ is 
transformed to the kinetic energy of the decay products instantaneously on 
the Hubble time scale. 

Thus, we have found that in both extreme cases the energy density in the 
throat initially scales like radiation if it is above the phase transition 
density. It is given by
\be
\label{ed}
\rho \, \sim \, g^{1/2} N^2_\UV \left( \frac{T_\RH}{M_4} \right)^3 T^4\,,
\ee
where $T$ is the standard model temperature. 
\begin{figure}[t]
\begin{center}
\includegraphics[scale=0.5]{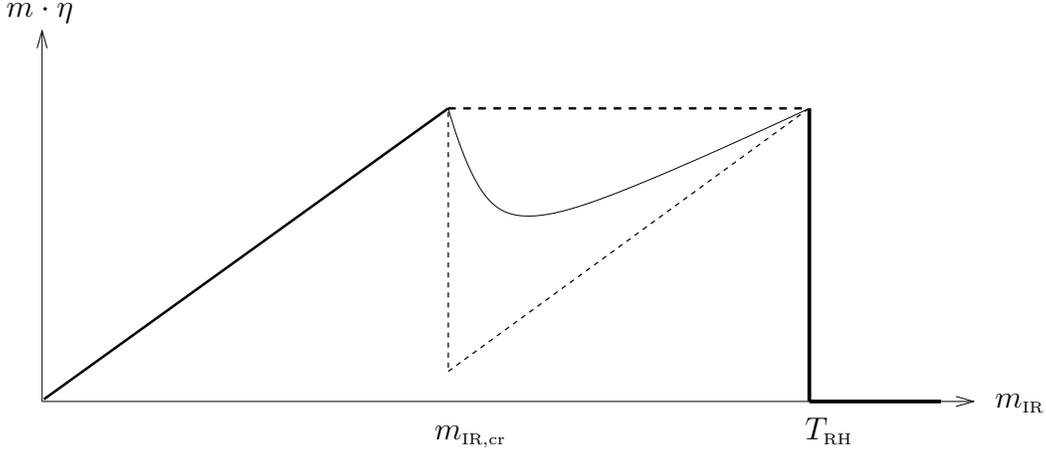} \put(-367,148){$m \cdot \eta$}
\put(7,0){$m_\IR$} 
\put(-205,-12){$m_\IRc$}
\put(-65,-12){$T_\RH$}
\vspace{.4cm}
\caption{Schematic plot of $m \cdot \eta$ as a function of the IR scale 
$m_\IR$ of the throat. Here, $m_\IRc $ is the IR scale for which $T_\pt \sim 
T_\RH$, i.e., for which the throat is heated precisely to its phase transition 
temperature. See text for more details.}
\label{fig:plot}
\end{center} 
\end{figure}
\noindent
We assume the scaling behaviour to change when the energy density has dropped 
to $\smash{\rho \sim N_\IR m^4_\IR}$. This happens when the standard model 
has a temperature
\be
\label{Tpt}
T_\pt \, \sim \, m_\IR \frac{ N_\IR^{1/4}}{ N_\UV^{1/2}}  \hspace{-1mm} 
\left( \frac{M_4}{T_\RH} \right)^{3/4} \hspace{-1mm},
\ee
where we have neglected a factor of $g^{1/8}$ which is close to 1. The energy 
density scales like matter afterwards and the ratio of energy density and 
entropy density, $ \rho/s = m \cdot \eta$, stays constant. Here $\eta=n/s$ 
is the glueball number density normalized by the entropy density. Using 
Eq.~\eqref{ed} and $\smash{s = g \frac{2 \pi^2}{45}  T^3}$ (dominated by the 
standard model sector) at $T=T_\pt$, we find the glueball mass density per
entropy density
\be
\label{energydensity2}
m \cdot \eta  \, \sim \,   \frac{N^2_\UV}{g^{1/2}} \left( \frac{T_\RH}{M_4} 
\right)^3 T_\pt.
\ee
The factor of $T_\pt$ is 
smaller than or equal to $T_\RH$ and reflects the fact that the corresponding 
energy density undergoes a phase of $\smash{a^{-4}}$ dilution. The quantity 
$m \cdot \eta$ is useful because it determines the contribution of the 
glueballs to the total energy density in the late universe. We have plotted 
$m\cdot \eta$ as a function of $m_\IR$ schematically in Fig.~\ref{fig:plot}. 
The part corresponding to Eq.~\eqref{energydensity2} is the straight bold line 
which grows linearly\footnote{
Note 
that this plot has to be read either at fixed $N_\UV$, in which case 
$N_\IR$ must be interpreted as function of $N_\UV$ and $m_\IR$, or at fixed 
$N_\IR$, in which case $N_\UV$ must be interpreted as function of $N_\IR$ and 
$m_\IR$. In both cases an extra logarithmic dependence of $m\cdot\eta$ on 
$m_\IR$ is introduced, which we however neglect. 
}
with the IR scale from $m_\IR = 0$ up to an $m_\IR$ such 
that $T_\pt$ in Eq.~\eqref{Tpt} is of the same order of magnitude as $T_\RH$. 
This is the maximal IR scale for which Eq.~\eqref{energydensity2} is valid 
because at this point the initial energy density in the throat, 
Eq.~\eqref{energydensity}, is of the same order of magnitude as the critical 
energy density $\rho \sim N_\IR m_\IR^4$.

Dividing Eq.~\eqref{energydensity2} by $\smash{\rho_c/s_0 \simeq 2 \cdot 
10^{-9}}$~GeV, where $\smash{\rho_c}$ is the current critical energy density 
for a flat universe and $s_0$ is the current entropy density, and using $g 
\sim 100$ as well as $M_4 \simeq 2 \cdot 10^{18}$ GeV, we have
\be
\label{Omega}
\Omega \, = \, \frac{\rho}{\rho_c} \,  \sim \, \left( \frac{T_\RH \,
N_\UV^{1/2}}{6 \cdot 10^{11} \, 
\text{GeV}} \right)^4 \, \left( \frac{T_\pt}{T_\RH} \right).
\ee
This is the contribution of the throat sector to the density parameter. The 
second factor is smaller than or equal to 1 and again reflects the fact that 
the corresponding energy density undergoes a phase of $\smash{a^{-4}}$ 
dilution.

Let us now consider the case that the energy density deposited in a throat, 
Eq.~\eqref{energydensity}, is smaller than $\rho \sim N_\IR m_\IR^4$. If the 
initial gauge theory state settles into a large number of slow glueballs, the 
energy density scales like matter from the beginning. Taking this scaling 
into account, the mass density over entropy density $m \cdot \eta$ is given 
by Eq.~\eqref{energydensity2} with $T_\pt$ replaced by $T_\RH$. Similarly, 
the contribution of the throat sector to the density parameter $\Omega$ is
given by Eq.~\eqref{Omega} with the second factor in brackets replaced by 1. As
a function of $m_\IR$, $m \cdot \eta$ is constant in this case, which we have 
plotted as the bold dashed line in Fig.~\ref{fig:plot}. 

The analysis is more subtle if the initial gauge theory state decays into 
an $\mathcal{O}(1)$ number of fast glueballs. The important question is 
again whether the glueballs thermalize because this determines the 
distribution of kinetic energies directly after reheating. As we will see in 
Sect.~\ref{subtleties}, in the cases of interest the glueballs do not 
thermalize if the initial energy density is smaller than $\rho \sim N_\IR 
m_\IR^4$. Since the glueballs are thus ultrarelativistic initially with 
kinetic energies of the order of $T_\RH$, the energy density scales like 
radiation until the glueballs become nonrelativistic. Taking this into 
account, the contribution of the throat sector to the total energy density is 
determined by Eqs.~\eqref{energydensity2} and \eqref{Omega} with $T_\pt$ 
replaced by $m_\IR$. We have plotted $m \cdot \eta$ as a function of $m_\IR$ 
for this case as the thin dashed line in Fig.~\ref{fig:plot}.

Finally, throats with $m_\IR > T_\RH$ are not heated for kinematic reasons. 
Therefore, the mass density over entropy density $m \cdot \eta$ is zero in 
this region, which we have plotted as a bold straight line in 
Fig.~\ref{fig:plot}. Now, the bold straight lines in Fig.~\ref{fig:plot} 
correspond to regions where the function is the same in both extremal cases. 
We therefore expect that, in these regions, the plotted function gives the 
true behaviour also in intermediate cases (e.g.~the decay of the initial 
gauge theory state into a large number of glueballs with a complicated 
distribution of kinetic energies). The thin dashed lines corresponds to the 
decay into a small number of highly-energetic glueballs whereas the bold 
dashed line corresponds to the decay into a large number of slow glueballs. 
We expect the true behaviour in this region to be in between these two 
extremes. We have plotted our expectation schematically as the thin curve.

\subsection{Subtleties with thermalization}
\label{subtleties}
In this section, we consider the question of thermalization for the case that 
the initial gauge theory state decays into an $\mathcal{O}(1)$ number of 
highly-energetic glueballs. Let us assume that this decay is fast\footnote{
As
already mentioned, the process corresponding to the primary vertex is always
fast since we are dealing with a strongly coupled system. However, similarly 
to QCD processes with final state jets, the hadronization time scale may be 
much slower. Thus, strictly speaking, we should derive the thermalization 
criterion taking into account the evolution of the decay products into 
glueballs. Since we are only interested in order-of-magnitude estimates, we
neglect these subtleties in the following. 
}
compared to 
the Hubble time scale Eq.~\eqref{rts}. The number density of the decay 
products is then given by Eq.~\eqref{nd} directly after reheating. If $n \, 
\langle \sigma v \rangle > H$, where $\langle \sigma v \rangle$ is the 
thermally averaged product of interaction cross section $\sigma$ and velocity 
$v$, the glueballs thermalize. Since the glueballs have high energies 
$E \sim T_\RH \gg m_\IR$, their velocity $v$ is close to $1$. For energies
\be
E \, \gg \, N_\IR^2 \, m_\IR, 
\ee
the scattering 
cross section of KK modes in a throat or, equivalently, of glueballs in the 
dual gauge theory fulfills the Froissart bound and is given 
by~\cite{Giddings:2002cd}
\be
\label{cs}
\sigma \, \sim m^{-2}_\IR \, \ln^2\left(\frac{E}{m_\IR}\right).
\ee
The glueball number density Eq.~\eqref{nd} scales as $\smash{n \propto 
a^{-3}}$. During radiation domination, $a \propto T^{-1}$. Therefore, $n 
\propto T^3$ as a function of the standard model temperature $T$. Since 
the Hubble rate Eq.~\eqref{rts} only scales as $\smash{H \propto T^2}$, the 
criterion for thermalization\footnote{The dependence 
of $\langle \sigma v \rangle$ on $T$ via the glueball energy $E$ is only 
logarithmic in the energy range of interest according to Eq.~\eqref{cs}.} is 
easiest to fulfill directly after reheating at $T \sim T_\RH$. Neglecting the 
logarithm in Eq.~\eqref{cs}, we find 
\be
\label{criterion}
m_\IR \, \lesssim N_\UV \frac{T^2_\RH}{M_4}
\ee
as a criterion for thermalization of a given throat sector with IR scale 
$m_\IR$. 

In Sect.~\ref{te}, we have assumed that the glueballs thermalize 
if and only if the initial energy density in the throat, 
Eq.~\eqref{energydensity}, is larger than or of the same order of magnitude 
as the critical energy density $\smash{\rho \sim N_\IR m_\IR^4}$. 
Using Eq.~\eqref{energydensity}, this criterion can be written as
\be
\label{criterion2}
m_\IR \, \lesssim \left( g^{-1/2} \, N_\IR \, N_\UV^2 \, \frac{T_\RH}{M_4}  
\right)^{-1/4} \; N_\UV \frac{T^2_\RH}{M_4}.
\ee
As we will discuss in Sect.~\ref{SingleThroat}, $N_\UV$ can vary in the range
$10 \lesssim N_\UV \lesssim 10^4$. It then follows from Eq.~(\ref{Omega}) that
the reheating temperature has to be between $10^{10}$~GeV and
$10^{11}$~GeV if throat dark matter is not to be negligible today. Given the
numerically large prefactor $g^{-1/2} N_\IR N_\UV^2$ in Eq.~(\ref{criterion2})
and the exponent $1/4$, it turns out that the criteria in
Eqs.~\eqref{criterion} and \eqref{criterion2} are roughly equivalent. This
justifies our previous simplifying assumption that a throat thermalizes if and
only if the energy deposited in it is higher than its critical energy density.

Finally, let us discuss throat sectors which thermalize according to 
Eq.~\eqref{criterion} in more detail. The high-energy scattering cross 
section of KK modes in Eq.~\eqref{cs} is dominated by the production of black 
holes localized near the IR end of the throat~\cite{Giddings:2002cd}.
In the gauge theory, these black holes correspond to so-called plasma-balls, 
localized lumps of gauge theory plasma above the critical temperature, which 
are classically stable~\cite{Aharony:2005bm}. We expect the thermalization 
to proceed as follows: Two glueballs incident with an impact parameter 
smaller than $\sqrt{\sigma}$ at high energy will produce a plasma-ball. Other 
glueballs will be absorbed by the plasma-ball~\cite{Aharony:2005bm} and 
different plasma-balls will merge into larger plasma-balls. If the energy 
density is large enough for the gauge theory to be in the deconfined phase, 
this process continues until the gauge theory plasma fills out the whole 
space. On the gravity side this corresponds to the growing of the black hole 
horizon until it completely replaces the IR end of the throat. Otherwise, if 
the energy density is such that the equilibrated gauge theory is in the 
confined phase, the plasma-balls hadronize again until the whole space is 
filled by a nonrelativistic glueball gas.

\section{Relics in a throat}
\label{candidates}

\subsection{Processes in the hidden sector}
\label{processes}

After a confinement phase transition, glueballs with mass of the order of the 
confinement scale and with different spin are formed. Similarly, if the gauge
theory does not thermalize, the initial gauge theory states created at 
reheating settle into a certain number of light glueballs. A number of 
papers~\cite{Krasnitz:2000ir,Caceres:2000qe,Amador:2004pz,Firouzjahi:2005qs,
Noguchi:2005ws,Dymarsky:2006hn,Berg:2006xy,Dymarsky:2007zs,Benna:2007mb} 
have calculated parts of the bosonic glueball spectrum of the KS gauge theory. 
In~\cite{Berg:2006xy}, masses of KK towers of $7$ coupled scalar fields and 
the 
graviton polarized parallel to the uncompactified dimensions were determined. 
In particular, several scalar states lighter than the lowest spin-2 state were 
found. Their numerical technique, however, does not reveal to which 
linear combinations of the $7$ scalar fields these masses belong. 
In~\cite{Caceres:2000qe}, the mass of the lowest KK mode of the dilaton was 
calculated using some approximations in the geometry. Again, it was found to 
be lighter than a spin-1 and a spin-2 state~\cite{Caceres:2000qe, 
Amador:2004pz}, but no other scalar states were calculated to compare with. 
In the light of these findings, we expect the lightest state in the bosonic 
sector to be a scalar glueball. We do not know, however, to which field 
fluctuations on the gravity side of the duality this glueball corresponds. 

The KS gauge theory has $\mathcal{N}=1$ supersymmetry and the lightest scalar 
glueball has a spin-$\smash{\frac{1}{2}}$ superpartner. In a 
phenomenologically viable setup, supersymmetry is broken and the masses of the 
scalar and the spin-$\smash{\frac{1}{2}}$ glueball are no longer degenerate. 
We have not completely settled the important question of the size of the mass 
splitting and which of the two superpartners is lighter. In the following, 
we will mainly be interested in scenarios with high-scale SUSY breaking. 
As we discuss in Sect.~\ref{extra}, we expect that the 
spin-$\smash{\frac{1}{2}}$ glueball is lighter than its scalar superpartner 
in this case. Nevertheless, we keep the discussion as general as
possible and allow for the two possibilities that the fermion
is lighter or heavier than the scalar.\footnote{
It may happen that the lightest fermionic glueball is not the superpartner of
the lightest bosonic glueball. The following discussion then stays correct if
one replaces the spin-$\smash{\frac{1}{2}}$ superpartner by this lightest
fermionic glueball. Moreover, it may happen that the mass of the lightest
bosonic glueball is larger than twice the mass of the lightest fermionic
glueball. The former could then decay to the latter via couplings discussed
below. We will not consider this possibility in the following.}

Generically, the glueball effective theory includes various cubic 
interactions. For example, for a scalar glueball $\mathcal{G}$, a 
spin-1 glueball $\smash{\mathcal{A}_\mu}$ and a spin-2 
glueball $\smash{\mathcal{H}_{\mu \nu}}$, there are couplings of the type 
\be
\label{cubiccouplings}
\partial_\mu \mathcal{G} \, \mathcal{A}_\nu \, \mathcal{H}^{\mu \nu}  + 
m_\IR \, \mathcal{A}_\mu \mathcal{A}^\mu \, \mathcal{G} + m_\IR^{-1} 
\partial_\mu \mathcal{G} \, \partial_\nu \mathcal{G} \, \mathcal{H}^{\mu \nu} 
+ \dots .
\ee
The coupling strengths follow on dimensional grounds up to possible factors
of $N_\IR$ that we have not determined. Also, there may be more partial 
derivatives involved or they may act differently. Due to interactions of this 
type, heavy glueballs decay quickly to a few light states which cannot decay 
further for kinematic reasons. Note, however, that the KS gauge theory has a
global SU(2)$ \times$SU(2) symmetry which forbids certain couplings of the
type of Eq.~\eqref{cubiccouplings}. From the dual gravity point of view, this
symmetry corresponds to an isometry of the KS throat. In a compactified setup,
the KS throat is attached to a Calabi-Yau manifold which breaks this isometry 
in the UV. This symmetry breaking is mediated to the IR as discussed 
in~\cite{Kofman:2005yz,Aharony:2005ez,Berndsen:2007my}. We therefore expect 
that couplings of the type of Eq.~\eqref{cubiccouplings}, which violate the
global symmetry, are nevertheless present, albeit with a possibly smaller
coupling strength. In the following, we ignore the effects of glueballs charged
under the SU(2)$ \times$SU(2) symmetry. In particular, we expect that the
lightest scalar glueball and its superpartner are singlets with respect to this
symmetry. 

If the gauge theory has thermalized, the glueballs which can not decay interact
with each other for a certain period of time after the confinement phase 
transition. This leads to a significant reduction of the abundances of all 
the states heavier than the lightest glueball, including its superpartner if 
the mass splitting from supersymmetry breaking is not too small. We now
analyse this effect in some detail:

For simplicity, we focus on only two glueball species. Generically, the 
glueball effective action includes couplings of the type
\be
\label{quarticcouplings}
\mathcal{H} \mathcal{H} \, \mathcal{G} \mathcal{G},
\ee
where $\mathcal{H}$ and $\mathcal{G}$ are the heavy and light glueball
respectively, and all Lorentz- and/or spinor-indices are appropriately 
contracted. By assumption, the masses of the two glueball species satisfy 
$\smash{m_\mathcal{G} < m_\mathcal{H}}$. As long as the two glueball species 
are in equilibrium, the density $n_\mathcal{H}$ of the heavy glueballs is 
suppressed relative to the light glueball density $n_\mathcal{G}$ by an 
exponential factor $\exp[-(m_\mathcal{H}-m_\mathcal{G})/\tilde{T}]$ after
the temperature $\tilde{T}$ of the glueball gas falls below $m_\IR$. This
exponential decrease of the number density of $\mathcal{H}$ glueballs
continues until they are so dilute that they decouple. This happens, when
\be
\label{freeze-out}
n_\mathcal{H} \cdot \langle \sigma v \rangle \, \sim \, H.
\ee
Here $\langle \sigma v \rangle$ is the thermally averaged product of cross 
section and relative velocity for the $2 \cdot \mathcal{H} \leftrightarrow 2 
\cdot \mathcal{G}$ process, which evaluates to~\cite{swo}\footnote{If the 
mass difference $m_\mathcal{H}  - m_\mathcal{G}$ is very small, this 
cross-section is kinematically suppressed. This may happen, for example, for 
the superpartner of the lightest glueball. These glueballs are then diluted 
to a lesser extent.}
\be
\langle \sigma v \rangle \, \sim \, m_\IR^{-2}.
\ee

Since $n_\mathcal{H}$ drops exponentially after the temperature $\tilde{T}$ 
falls below $m_\IR$, the heavy glueballs decouple when the temperature of the 
glueball gas is still of the same order of magnitude as the IR scale. We can 
therefore derive the freezeout density of the heavy glueballs from 
Eq.~(\ref{freeze-out}) using the phase-transition Hubble rate $H(T_\pt)$. 
Furthermore, we can approximate the light glueball density by $m_\IR^3$. The 
ratio of heavy and light glueball densities directly after freezeout, i.e. the 
dilution factor, is then given by 
\be
\label{dilutionfactor}
\frac{n_\mathcal{H}}{n_\mathcal{G}} \, \sim \, \frac{H(T_\pt)}{m_\IR} \, 
\sim \, \frac{g^{1/2} m_\IR M_4^{1/2}}{ N_\UV T_\RH^{3/2}}\,.
\ee
Here we have calculated the Hubble rate according to Eqs.~(\ref{rts}) and 
(\ref{Tpt}) and disregarded a small power of $N_\IR$. Our formula is valid 
if the right-hand side is smaller than 1. If, however, the right-hand side 
is formally larger than 1, the $\mathcal{H}$ glueballs are decoupled from the 
beginning and not diluted at all.

\begin{figure}[t]
\begin{center}
\begin{minipage}{6.2cm}
\begin{center}
\includegraphics[scale=0.5]{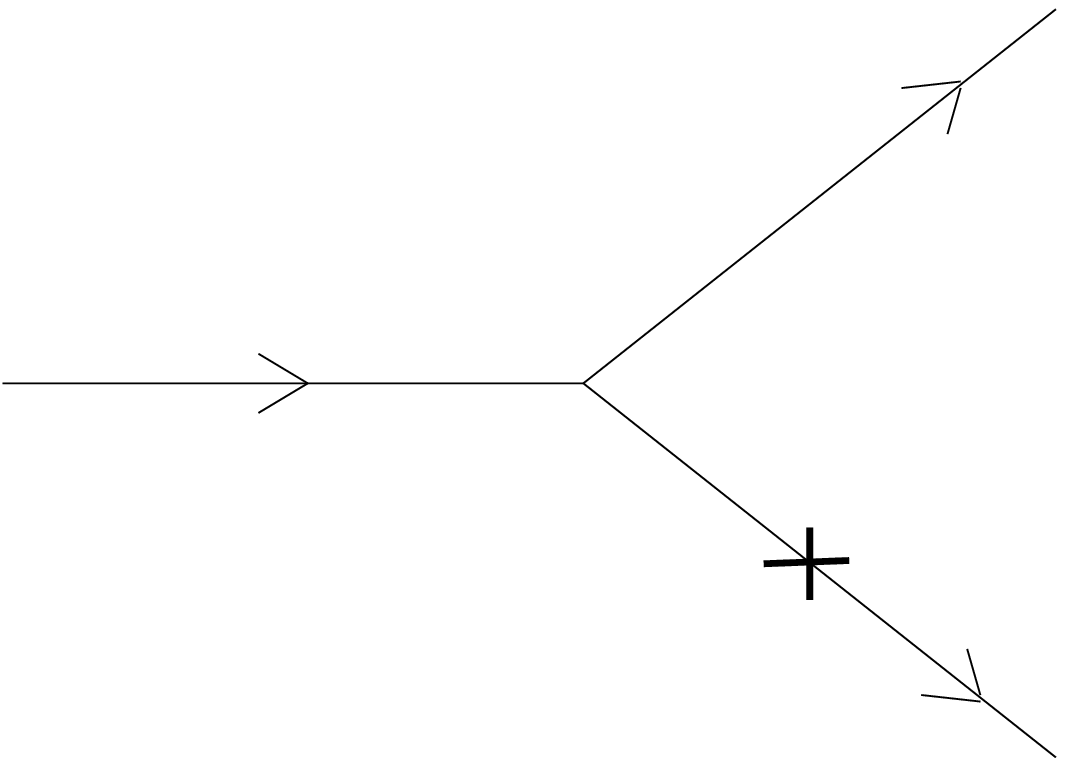}
\put(-23,-3){$h_{\mu \nu}$}
\put(-17,104){$\mathcal{G}$} 
\put(-67,31){$\mathcal{H}_{\mu \nu}$}
\put(-140,60){$\mathcal{B}$}
\caption{Decay of a bosonic glueball $\mathcal{B}$ into a scalar glueball 
$\mathcal{G}$ and a graviton $h_{\mu \nu}$ via a spin-2 glueball 
$\mathcal{H}_{\mu \nu}$.\label{fig:decay1}}
\end{center}
\end{minipage}
\hspace{1.4cm}
\begin{minipage}{6.2cm}
\begin{center}
\vspace{-.6cm}
\includegraphics[scale=0.5]{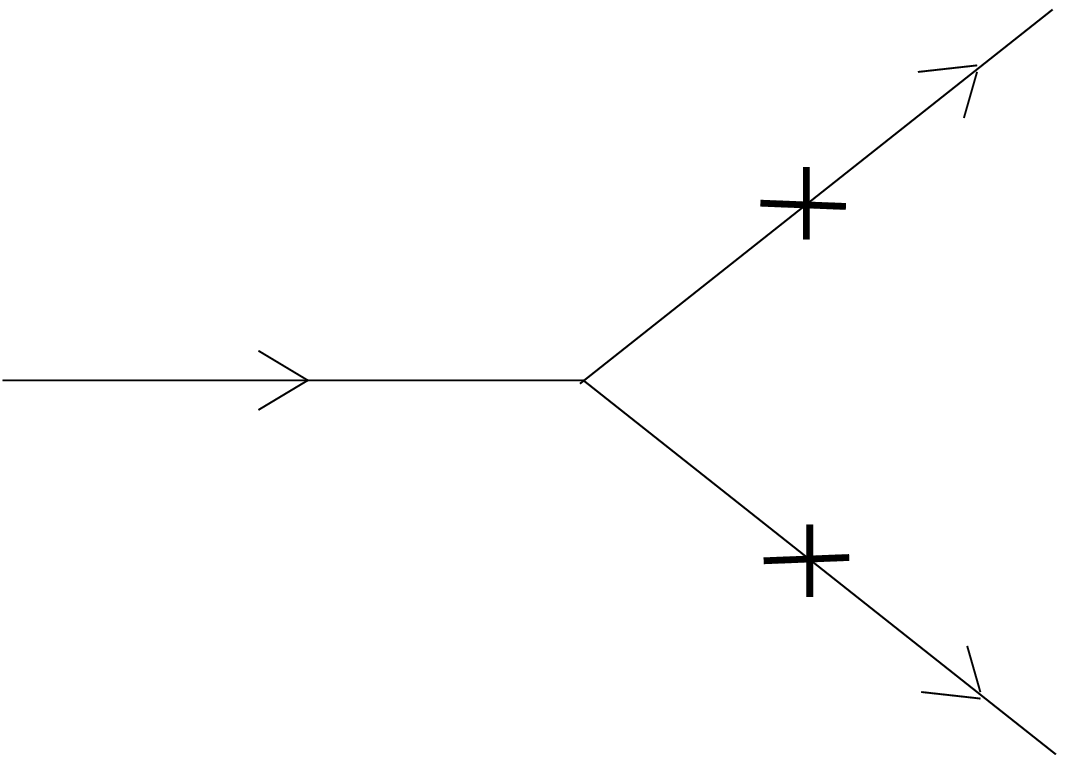}  
\put(-23,-3){$h_{\mu \nu}$} 
\put(-23,108){$h_{\mu \nu}$} 
\put(-67,75){$\mathcal{H}_{\mu \nu}$}
\put(-67,31){$\mathcal{H}_{\mu \nu}$}
\put(-140,60){$\mathcal{G}$}
\caption{Decay of a scalar glueball $\mathcal{G}$ into two 
gravitons $h_{\mu \nu}$ via spin-2 glueballs $\mathcal{H}_{\mu \nu}$.
\label{fig:decay2}}
\end{center}
\end{minipage}
\end{center}
\end{figure}

At a later time, all bosonic glueballs which are left over from this 
process decay by emission of a graviton into the lightest scalar glueball. 
Namely, as we will discuss in Sect.~\ref{DecaytoSM}, spin-2 glueballs mix 
with the 4d graviton. The corresponding vertex will be derived in 
Sect.~\ref{DecaytoSM} 
and is given in Eq.~\eqref{vertex2}. Combined with interactions of the type 
given in Eq.~\eqref{cubiccouplings}, processes such as that shown in 
Fig.~\ref{fig:decay1} are possible: A bosonic glueball $\mathcal{B}$ 
decays to the lightest scalar glueball $\mathcal{G}$ and, via a virtual 
spin-2 glueball $\mathcal{H}_{\mu \nu}$, to the graviton $h_{\mu \nu}$. 
Since the mixing between graviton and spin-2 glueball is $\sim N_\UV/M_4$
(see Eq.~\eqref{vertex2}) and $m_\IR$ is the only other relevant
dimensionful parameter, we have
\be
\label{dr1}
\Gamma \, \sim \, N_\UV^2 \, \frac{m_\IR^3}{M_4^2}
\ee
for the corresponding decay rate (up to an unknown factor related to 
$N_\IR$ which may result from the three-glueball vertex). Similarly, 
fermionic glueballs decay to the lightest spin-$\smash{\frac{1}{2}}$ glueball 
and a graviton. The corresponding decay rate is again given by 
Eq.~\eqref{dr1}. In addition, fermionic glueballs may in principle decay to 
lighter bosonic glueballs and vice versa by emission of a gravitino. Indeed, 
by supersymmetry we expect a mixing of spin-$\smash{\frac{3}{2}}$ glueballs 
with the gravitino, again with the vertex given in Eq.~\eqref{vertex2}. This 
would allow for the corresponding processes. As we have mentioned at the 
beginning of this section, we mainly focus on a setup in which the gravitino 
is very heavy and such decays are kinematically forbidden. In this case, a 
non-negligible amount of the superparter of the lightest glueball is left.

At an even later time, the lightest scalar glueballs decay into two gravitons. 
Indeed, using the three-glueball coupling and the vertex in 
Eq.~\eqref{vertex2}, the process shown in Fig.~\ref{fig:decay2} is possible. 
The corresponding decay rate is
\be
\label{dr2}
\Gamma \, \sim \, N_\UV^4 \, \frac{m_\IR^5}{M_4^4}
\ee
up to an unknown factor depending on $N_\IR$. By supersymmetry, we expect 
that the lightest spin-$\smash{\frac{1}{2}}$ glueballs may decay to a 
graviton and a gravitino with the same rate. As before, if the gravitino is 
very heavy, such decays are kinematically forbidden.

\subsection{Decay to the standard model sector}
\label{DecaytoSM}

Of the processes described so far, only the decay to two gravitons (or to
graviton and gravitino, if the gravitino is light enough) can be sufficiently 
slow to be relevant for late cosmology. The other processes have a time
scale much shorter than the age of the universe for all relevant choices of 
parameters. Thus, in late cosmology, the energy density in the throat sector 
is completely in the form of the lightest glueball and its superpartner. 
The abundance of the heavier superpartner is depleted by the factor in 
Eq.~\eqref{dilutionfactor} if the gauge theory was thermalized. 

The glueballs couple (very weakly) to the standard model and other throats. 
Therefore, they may decay to these sectors. In~\cite{hotthroats}, we 
have calculated the decay rate of a dilaton between two AdS$_5 \times$S$^5$ 
throats embedded in a 6d torus. Again, we chose the dual picture where the 
throats are replaced by corresponding brane stacks. The calculation is then 
almost identical to the calculation of the energy transfer rate described at 
the beginning of Sect.~\ref{thermalproduction}. A crucial difference is the 
fact that the coupling between the glueball on a brane stack (corresponding 
to the dilaton in the throat) and the dilaton in the embedding torus cannot 
be read off from any Lagrangian since the glueball is a non-perturbative 
object. Therefore, in~\cite{hotthroats}, we first used the gravity picture to 
calculate the decay rate for the following simplified setup: We considered a 
single finite throat embedded in flat 10d space and a dilaton KK mode 
decaying from the throat to the asymptotically flat region. Then, we 
determined the glueball-dilaton vertex from the requirement that the decay 
rate be reproduced in the gauge theory picture. Let us restrict ourselves to 
a glueball corresponding to an s-wave of the dilaton with respect to the 
S$^5$ in the throat. The vertex between the dilaton in the embedding torus 
and such a glueball on a brane stack is then given by (cf.~Eq.~(37) of~\cite{
hotthroats})
\be
\label{vertex}
N_\UV \frac{m_\IR^{1/2} m^{5/2}}{M_{10}^4}\,,
\ee
where $m$ is the mass of the glueball (for our purposes $m\sim m_\IR$). Note 
that Eq.~(\ref{vertex}) gives the vertex in the 10d effective theory. Thus, 
to be part of a Lagrangian, this vertex has to be multiplied by a 6d 
$\delta$-function for the brane stack, the 4d glueball field of mass 
dimension 1 and the 10d dilaton field of mass dimension 4. As we have 
outlined in~\cite{hotthroats}, Eq.~(\ref{vertex}) also works for a KS 
throat as long as the field corresponding to the glueball satisfies the 
equations of motion of a massless 5d scalar. 

A simple example is provided by the 10d graviton $\hat{h}_{\mu \nu}$ 
polarized parallel to the uncompactified dimensions which fulfills the 
equation of motion of a massless, minimally coupled 10d scalar field~\cite{
Csaki:2000fc,Firouzjahi:2005qs}. It also obeys the equations of motion of 
a massless scalar in the 5d effective theory of the throat. The vertex 
between the corresponding glueballs and 
$\hat{h}_{\mu \nu}$ in the embedding manifold is again given by 
Eq.~\eqref{vertex}. We want to calculate the decay rate of such glueballs to 
another brane stack on which the standard model may live or which may 
correspond to another throat. This decay is mediated by the tower of KK 
modes of $\hat{h}_{\mu \nu}$ in the embedding manifold. In analogy to the 
energy transfer rate of 
Sect.~\ref{thermalproduction}, the contribution of the zero mode in this KK 
expansion dominates if the distance $A$ between the two brane stacks is of 
the same order of magnitude as the size $L$ of the embedding 
manifold.\footnote{
More 
generally, there is again a term with an $A^{-8}$ dependence in the 
expression for the decay rate. This term becomes dominant at small $A$, 
see~\cite{hotthroats}.
} 
Normalizing this zero mode contributes an extra factor of $L^{-3}$ to the 
vertex of Eq.~(\ref{vertex}). Using $\smash{M_4 \simeq M_{10}^4 L^3}$, the 
vertex of the 4d effective theory characterizing the mixing of the spin-2 
glueball and the 4d graviton is thus given by
\be
\label{vertex2}
N_\UV \frac{m_\IR^{1/2} m^{5/2}}{M_4}\,.
\ee
This is similar to the mixing between the photon and the $\rho$ meson known 
from QCD and was also observed in~\cite{Batell:2007jv} for the gauge theory 
dual of a 5d RS model. Using Eq.~\eqref{vertex2} and the fact that the 4d 
graviton couples to the energy-momentum tensor on the other brane stack with 
strength $\smash{M_4^{-1}}$ as well as summing over all degrees of freedom 
on the other brane stack, we get
\be
\label{drSpin2}
\Gamma \, \sim \, N^2 \, N^2_{\UV} \frac{m^4 m_\IR}{M_4^4}.
\ee
This is the decay rate of a spin-2 glueball (which is a singlet with respect 
to the R-symmetry of the gauge theory) to another brane stack. 

The vertex Eq.~\eqref{vertex2} also applies to the coupling of scalar 
glueballs to zero modes of other fields in the embedding manifold. To 
see this, let us consider the dilaton $\phi$, whose equation of motion is
\be
\label{dilatonEOM}
\nabla^2 \phi \, = \, \frac{1}{12} \,e^\phi \, \tilde{F}_{MNP} \tilde{F}^{MNP} 
- \frac{1}{12} \, e^{- \phi} H_{MNP} H^{MNP}.
\ee 
Here, $\tilde{F_3}= F_3 - C H_3$ and $F_3 = d C_2$ and $H_3 = d B_2$ are the 
field strengths of the Ramond-Ramond 2-form $C_2$ and the Neveu-Schwarz 2-form 
$B_2$, respectively. Moreover, $C$ is the Ramond-Ramond scalar, which we have 
taken to be constant in Eq.~\eqref{dilatonEOM}. In a background with imaginary 
self-dual 3-form flux~\cite{Giddings:2001yu}, the flux fulfills 
\be
\label{ISDflux}
H_{MNP} H^{MNP} = e^{2 \phi} \, \tilde{F}_{MNP} \tilde{F}^{MNP}
\ee
for the background value of $\phi$ and the right-hand side of 
Eq.~\eqref{dilatonEOM} vanishes. This is no longer the case if one shifts 
the background value of $\phi$ while keeping $B_2$ and $C_2$ fixed. 
However, if one simultaneously shifts $B_2$ in such a way that 
Eq.~\eqref{ISDflux} remains fulfilled, the right-hand side of 
Eq.~\eqref{dilatonEOM} still vanishes. In other words, there exists a flat 
direction in the 5d effective theory which one can parameterize, e.g., by 
the value of the dilaton. The corresponding field then fulfills the equation 
of motion of a 5d minimally coupled, massless scalar. A light glueball, i.e. 
a KK mode localized at the bottom of the throat, will generically mix with 
this flat direction in the upper part of the throat~\cite{Giddings:2005ff, 
Frey:2006wv}. Thus, scalar glueballs couple to zero modes in the embedding 
manifold with the previously derived vertex Eq.~\eqref{vertex2}. 

Note also that stronger couplings may arise for glueballs mixing with 
fields of the 5d effective theory which have tachyonic mass. The reason is 
that the 5d profile of such fields is suppressed more weakly if one moves 
from the IR to the UV end of the approximate AdS$_5$ geometry. It may 
be worthwhile to investigate this effect in more detail in the future. 

A crucial difference to the decay of spin-2 glueballs is the fact that the 
mediating field is massive if fluxes are present in the compact 
space containing the two brane stacks~\cite{Giddings:2001yu}. More 
specifically, the zero mode of the (axion-)dilaton and the complex structure 
moduli get a mass 
\be
\label{mtau}
m_\tau \, \sim \,  \frac{M_{10}^2}{M_4}, 
\ee
where we have assumed that $g_s \sim 1$. The K\"ahler moduli can be lighter. We
discuss the possible effects of K\"ahler moduli in mediating decays in 
Sect.~\ref{extra}. 

Redoing the calculation leading to 
Eq.~\eqref{drSpin2} with a massive instead of a massless propagator for the 
mediating field, we get an extra factor of 
\be
\label{suppressionfactor}
\left( \frac{m^2}{m^2-m_\tau^2} \right)^2 \, \sim  \,
\left( \frac{m}{m_\tau} \right)^4
\ee
for the decay rate of a scalar glueball. In the last step, we have assumed
that the dilaton (or the complex structure modulus) is heavier than the
decaying glueball. In this case, Eq.~\eqref{suppressionfactor} suppresses the
decay rate.\footnote{
One can check that the 
contribution of higher KK modes is still smaller than the zero mode 
contribution for $m_\tau$ given in Eq.~\eqref{mtau}.} 
Combining Eqs.~\eqref{drSpin2} and \eqref{suppressionfactor} and using
Eq.~\eqref{mtau}, we then get
\be
\label{scalardr}
\Gamma \, \sim \, N^2 \, N^2_{\UV} \frac{m^8 m_\IR}{M_{10}^8}
\ee
for the decay rate of a scalar glueball to another brane stack. In particular,
for $\smash{N^2 = g}$, Eq.~\eqref{scalardr} is the decay rate of scalar 
glueballs to the standard model.

We have not yet determined the decay rate of the spin-$\smash{\frac{1}{2}}$ 
glueballs. For unbroken supersymmetry, this rate is related to the
scalar decay rate by a supersymmetry transformation and both rates are thus 
equal. We expect that even for broken supersymmetry, the relevant vertices 
agree up to $\mathcal{O}(1)$ prefactors and that the suppression of the 
spin-$\smash{\frac{1}{2}}$ decay rate by the dilatino propagator 
is the same as the suppression of the scalar decay rate by the dilaton 
propagator. Therefore, we can use Eq.~\eqref{scalardr} also for the decay rate
of spin-$\smash{\frac{1}{2}}$ glueballs to other throats. 

The situation is more complicated for decays to the standard model. 
Namely, we have assumed that the gravitino is much heavier than the glueballs
and, accordingly, that supersymmetry is broken at a high scale.
This means that also the superpartners of standard model particles are heavier
than the decaying spin-$\smash{\frac{1}{2}}$ glueballs. If R-parity is
conserved, most decay channels involve such a superpartner as a final state and
the corresponding decays are therefore kinematically forbidden. A coupling that
does not involve a superpartner is
\be
\label{vertex3}
\lambda \, \bar{l} \, \psi \, H,
\ee
where $l$ is a lepton doublet, $H$ is the Higgs doublet and $\psi$ is a 
dilatino or any other modulino.\footnote{This coupling was already considered
in~\cite{Benakli:1997iu} since it also leads to a mixing between the modulino
and the neutrino.} The coupling strength $\lambda$ may be $\mathcal{O}(1)$ or 
it may be suppressed as $\lambda = m /M_4$ where $m$ is some
low mass scale. The coupling in Eq.~\eqref{vertex3} probably requires R-parity
violation. Namely, the corresponding coupling containing the modulus instead of
the modulino generates a bilinear R-parity violating term for nonzero modulus
vev. But since we assume high-scale supersymmetry breaking, a large coupling
$\lambda$ may still be allowed. Moreover, even for maximally broken R-parity,
all other decay channels involve standard model superpartners which further
decay into standard model particles. The corresponding decay rates are 
therefore suppressed by the propagators of the heavy superpartners and are 
smaller than the decay rate resulting from Eq.~\eqref{vertex3}. Therefore, we
concentrate on this coupling in the following. Redoing the steps leading to
Eq.~\eqref{scalardr} with the vertex of Eq.~\eqref{vertex3}, we find
\be
\label{fermionicdr}
\Gamma \, \sim \, \lambda^2 \, N^2_{\UV} \frac{m^6 m_\IR}{M_{10}^8 /M_4^2}
\ee
for the decay rate of spin-$\smash{\frac{1}{2}}$ glueballs to the standard
model.  If the coupling in
Eq.~\eqref{vertex3} is absent and R-parity is exactly
conserved, the spin-$\smash{\frac{1}{2}}$ glueballs can not decay at all to the
standard model sector. If, in addition, there is no throat with lower IR scale
(otherwise decays to this sector with the rate in Eq.~\eqref{scalardr} are
possible), the spin-$\smash{\frac{1}{2}}$ glueballs are absolutely stable. 

Finally, the gravitino may also play a role in mediating decays of fermionic 
glueballs. We will discuss these effects in Sect.~\ref{extra}.

\section{Cosmological scenarios}
\label{scenarios}
If the glueballs from a given throat are stable until our epoch, then 
these glueballs are an interesting dark matter candidate. We begin our 
discussion in 
Sect.~\ref{SingleThroat} with scenarios where a single throat accounts for the 
observed dark matter. Essentially, this gives a relation for the required 
reheating temperature as a function of the IR scale of the throat. In
Sect.~\ref{ManyThroats} we discuss the probably more natural scenario that
various throats of different lengths are present. As we show, in this case a
reheating temperature of $\smash{10^{10} - 10^{11}} $~GeV naturally leads to 
the right dark matter abundance.

\subsection{A single throat}
\label{SingleThroat}

We consider a setup in which the compact manifold contains a 
single throat. In order to evaluate the relevant equations from
Sects.~\ref{thermalproduction} and \ref{candidates}, we have to fix $N_\IR$ and
$N_\UV$. These numbers determine the warp factor $h= \exp( 2 \pi N_\UV / 3
N_\IR)$ which in turn is related to the IR scale $\smash{m_\IR \sim h^{-1}
N_\IR^{-1/4} M_{10}}$ of the throat. To simplify the discussion and to avoid
uncertainties associated with unknown factors of $N_\IR$ in the various
glueball decay rates, we focus on throats where 
$N_\IR=\mathcal{O}(1)$. $N_\UV$ is then a function of $m_\IR$ and $M_{10}$. 
For our purposes it will be sufficient to use the typical values $N_\UV\sim 
10$ for long throats (e.g. for $m_\IR\sim 10^6$~GeV and $M_{10}\sim 
10^{15}$~GeV) and $N_\UV \sim 4$ for shorter throats (e.g. for $m_\IR\sim
10^{11}$~GeV). 

As we have explained in Sect.~\ref{te} and as one can see from 
Fig.~\ref{fig:plot}, we expect the glueball mass density over entropy density 
$m \cdot \eta$ or, equivalently, the contribution of glueballs to the density
parameter $\Omega$ to be maximized for throats with $m_\IR\sim m_\IRc$ and
with $m_\IR\sim T_\RH$. Let us focus on throats with the former IR scale
first. This IR scale is defined by the condition that the initial energy
density in the throat, Eq.~\eqref{energydensity}, is just the critical energy
density $\smash{\rho \sim N_\IR m_\IR^4}$ of that throat. Solving for $m_\IR$,
we have
\be
\label{mass}
m_\IRc \, \sim \, \left( \frac{  N_\UV^2 \, T_\RH^7}{N_\IR \, M_4^3} 
\right)^{1/4},
\ee
where we have neglected a factor of $\smash{g^{1/8}}$ which is 
$\mathcal{O}(1)$. It follows from Eq.~\eqref{Tpt} that $T_\pt \sim
T_\RH$ for this IR scale. Using $N_\UV \sim 10$ and $T_\pt \sim T_\RH$ in
Eq.~\eqref{Omega}, we see that a reheating temperature of the order of
$\smash{ 10^{11}}$
GeV leads to the right amount of dark matter. It follows from Eq.~\eqref{mass}
that the mass of this dark matter candidate is
\be
\label{massglueball}
m_\IR \, \sim \, 10^6   \text{ GeV}.
\ee

If the coupling in Eq.~\eqref{vertex3} is present, the 
spin-$\smash{\frac{1}{2}}$ glueballs decay to the standard model with a
rate given by Eq.~\eqref{fermionicdr}. The resulting lifetime is
\be
\label{lt1}
\tau \, \sim \, 10^{26} \left(  \frac{  M_{10} \cdot \lambda^{-1/4}}{2 \cdot 
10^{16} \, \text{GeV}} \right)^8 \text{ s}.
\ee
Note that we consider a setup in which the gravitino is
heavier than the glueballs. The decay of spin-$\smash{\frac{1}{2}}$ glueballs
to a graviton and a gravitino is therefore kinematically forbidden. It is not
sufficient to make the lifetime in Eq.~\eqref{lt1} just longer than
the present age of the universe (which is $\sim 10^{17}$~s): The glueball
decays produce
photons (e.g.~via hadronic showers) which contribute with a continuous
spectrum to the diffuse $\gamma$-radiation. The $\gamma$-ray flux
measured e.g.~by the experiment EGRET gives constraints on the lifetime of
unstable particles in dependence of their
mass density~\cite{KribsRothstein}.\footnote{See
e.g.~\cite{UP1,Bertone:2007aw,DecayingDarkMatter} for other work on decaying
dark matter.} In particular, an unstable particle with the 
mass density of dark matter has to live longer than $\sim 10^{26}$~s to comply
with
observations.\footnote{Here, we have used that the hadronic branching ratio
for decays via the coupling in Eq.~\eqref{vertex3} is $\mathcal{O}(1)$. For
decays exclusively to photons or leptons, the constraints are less severe.}
Thus, it depends on the two unknown parameters $M_{10}$ and $\lambda$ whether
the spin-$\smash{\frac{1}{2}}$ glueballs are a good dark matter candidate or
not.

An interesting scenario is a setup in which
$\lambda=\mathcal{O}(1)$.\footnote{
The coupling in
Eq.~\eqref{vertex3} leads to a mixing between the modulino $\psi$ and a
(left-handed) neutrino $\nu$ after electroweak symmetry breaking. Since the
modulino has a large mass $m_\tau$, the seesaw mechanism results in a light
mass eigenstate. For $M_{10} \sim 10^{16}$ GeV (the minimal value allowed
for $\lambda \sim 1$ according to Eq.~\eqref{lt1}), Eq.~\eqref{mtau} gives
$m_\tau \sim 10^{14}$ GeV. Using this value and the mixing mass term for
$\lambda \sim 1$ in the seesaw formula, the resulting neutrino mass is $\sim
0.1$ eV. Interestingly, this is precisely the mass range indicated by various
experiments.
}  
To get
a viable dark matter candidate, the 10d Planck scale has a rather limited
range in this case according to Eq.~\eqref{lt1}. This makes it more probable
that the lifetime of the glueballs is in a range that can be probed by more
sensitive $\gamma$-ray telescopes like the upcoming satellite GLAST.
If this scenario with $\lambda=\mathcal{O}(1)$ is realized in nature, one may
be able to see a signal in the near future.

The scalar glueballs from a throat with IR scale $ 10^6$~GeV decay to two
gravitons after $ 10^{15}$~s. Note, however, that this lifetime is
proportional to $\smash{m_\IR^{-5}}$ (cf.~Eq.~\eqref{dr2}). The
lifetime will
thus be somewhat larger or smaller for IR scales slightly different from
$\smash{10^6}$~GeV. If the lifetime is in the range of $10^{17}$~s (the
present age of the universe) to $10^{12}$~s (the time of matter-radiation
equality), the resulting decrease in the dark matter density may have
interesting consequences for structure formation. As we will explain in
Sect.~\ref{extra}, we expect the fermionic glueballs to be lighter than their
scalar superpartners if the supersymmetry breaking scale is larger than the IR
scale of the throat. Due to the mass difference, part of the scalar glueballs
annihilate into their superpartners after the phase transition and the
abundance of the scalar glueballs is depleted by the factor given in
Eq.~\eqref{dilutionfactor}. Inserting the above values for $N_\UV$, $m_\IR$
and $T_\RH$ in Eq.~\eqref{dilutionfactor}, we see that the scalar glueballs
make up for only $\smash{10^{-2}}$ of the total dark matter abundance and the
loss of mass density by the decay of the scalar glueballs is correspondingly
small. We can not exclude the possibility, however, that either the mass
splitting due to supersymmetry breaking is very small or that the fermionic
glueballs are heavier than a scalar superpartner. In these cases, the
scalar glueballs are not diluted and the loss of
dark matter mass density by the decay of the scalar glueballs is much larger.
Namely, it drops by a factor of $\smash{\frac{1}{2}}$ in the former case and
by a factor of $\smash{10^{-2}}$ in the latter case. It would be interesting
to analyse these possibilities and their implications for cosmology in more
detail. To this end, a better understanding of the effect of supersymmetry
breaking on the glueball mass spectrum would be required.

If the lifetime of scalar glueballs is large enough, a non-negligible
amount still exists at our epoch. Their decays to the standard model again
produce $\gamma$-radiation. Using Eq.~\eqref{scalardr} with $m_\IR \sim
10^6$~GeV, the partial lifetime of scalar glueballs for
decays to the standard model is
\be
\label{lt5}
\tau \, \sim \, 10^{26} \, \left( \frac{M_{10}}{3 \cdot 10^{13} \text{ GeV}} 
\right)^8 \text{ s}.
\ee
If the scalar glueballs (still) make up an $\mathcal{O}(1)$ fraction of the
dark matter at our epoch, this partial lifetime has to be larger than
$\smash{\sim 10^{26}}$~s to comply with the EGRET measurements.
If the current abundance of scalar glueballs is reduced (by decays to two
gravitons or by annihilation if the fermionic superpartner is lighter), the
lower bound on the lifetime becomes correspondingly
weaker. 

In contrast to fermionic glueballs, scalar glueballs can decay directly to
two photons. Decays via this channel in the halo of our galaxy lead to a sharp
$\gamma$-line in addition to the continuous spectrum. The $\gamma$-rays at 
energies around $\smash{ 10^6}$~GeV cannot be measured by EGRET or GLAST.
Ground-based $\gamma$-ray telescopes like HESS have the necessary energy
range, but a limited sensitivity due to the cosmic-ray background. At $10^6$
GeV, the measured flux in the cosmic ray spectrum is (see
e.g.~\cite{Horandel:2004bd})
\be
F \, \sim \,   10^{-12} \, ( \text{m}^2 \text{ sr s GeV})^{-1}.
\ee
To be detectable against this background, the flux from the decaying glueballs 
in the halo has to be of the same order of magnitude. This flux is emitted as a
sharp line at energy $ \sim m_\IR$ but smeared out by the detector due to a
finite energy resolution $\Delta E$. We model this effect by replacing the
$\delta$-function peak of the flux by a box of width $\Delta E$. The flux is
also inversely proportional to the mass $m_\IR$ and the lifetime $\tau$ of the
glueballs. Assuming that the scalar glueballs make up an $\mathcal{O}(1)$
fraction of the dark matter at our epoch, it is given by (see
e.g.~\cite{Bertone:2007aw})
\be
F \, \sim \,  10^{-12} \, \left( \frac{10^5 \text{ GeV}}{\Delta E}
\right) \, \left( \frac{10^6 \text{ GeV}}{m_\IR} \right) \, \left( 
\frac{10^{26}\text{ s}}{\tau} \right)\, ( \text{m}^2 \text{ sr s GeV})^{-1}.
\ee
For $m_\IR \sim 10^6$~GeV and $\Delta E \sim 10^{-1} \cdot E \sim 10^5$~GeV
as quoted by the HESS collaboration, the partial lifetime of the scalar
glueballs (for decays to the standard model) has to be less than
$\sim 10^{26}$~s to be detectable against the cosmic ray background. 
If the partial lifetime is somewhat larger, the $\gamma$-line may nevertheless
become detectable in the near future with an improved rejection of cosmic ray
events and a better sensitivity and energy resolution.

Thus, if an $\mathcal{O}(1)$ fraction of the dark matter at our epoch are
scalar glueballs and if their partial lifetime is not much larger than
$ 10^{26}$ s, two experiments may see a signal: The contribution of glueball
decays to the $\gamma$-ray spectrum below $10^2$~GeV may be detected by
GLAST. Furthermore, the $\gamma$-line near $10^6$~GeV may be
seen by HESS. A lifetime of the order of $ 10^{26}$~s follows if $M_{10} \sim
10^{13}$~GeV
according to Eq.~\eqref{lt5}. Such a low 10d Planck scale may be realized in a
large-volume compactification along the lines of~\cite{largevolume}. Note that
this scenario is incompatible with the aforementioned scenario in
which $\lambda=\mathcal{O}(1)$: According to Eq.~\eqref{lt1}, $\lambda$ has to
be very small (or zero) for such a low 10d Planck scale.

We can also discuss throats with IR scales smaller than $\smash{10^6}$
GeV. As before, we take $N_\UV\sim10$. According to
Eqs.~\eqref{Tpt} and \eqref{Omega}, $\Omega$ is proportional to
$\smash{T_\RH^{9/4} m_\IR}$. In order still to have the abundance of dark
matter with $\Omega \sim 1$, we have to increase the reheating temperature as
$\smash{T_\RH \propto m_\IR^{-4/9}}$ if we lower the IR scale. For example,
for a throat with IR scale $\smash{10^4}$~GeV, a reheating temperature of
$\smash{10^{12}}$~GeV would give the right abundance. Since the
various glueball decay rates are proportional to $m_\IR$ to some positive
power, the glueballs become more stable for lower IR scales.

Let us now consider throats with $m_\IR \sim T_\RH$ where $m \cdot \eta$ has
another peak. We take $N_\UV \sim 4$ in order to have
$N_\IR = \mathcal{O}(1)$. According to Eq.~\eqref{Omega} for $T_\pt \sim 
T_\RH$, such a throat again gives the right amount of dark matter for a
reheating temperature $\smash{ \sim 10^{11}}$~GeV. The mass of this dark
matter candidate correspondingly is $\smash{ \sim 10^{11}}$~GeV.
We expect that the glueballs are never in thermal equilibrium for such
short throats. Therefore, the heavier superpartners do not annihilate into the
lightest glueball states and the initial abundance of scalar and
spin-$\smash{\frac{1}{2}}$ glueballs is equal. The scalar glueballs decay to
two gravitons already after $\smash{10^{-8}}$ s, according to
Eq.~\eqref{dr2}. If the coupling in Eq.~\eqref{vertex3} is present, the
spin-$\smash{\frac{1}{2}}$ glueballs decay to the standard model after 
\be
\label{llt1}
\tau \, \sim \, 10^{27} \left(  \frac{  M_{10} \cdot \lambda^{-1/4} }{ 5
\cdot 
10^{20} \text{ GeV}}\right)^8 \text{ s}. 
\ee
Hadronic decays of particles in this mass range have been considered 
in~\cite{Birkel:1998nx} to explain events in the cosmic ray spectrum
beyond the GZK cutoff. Taking the measured flux in this energy range, claimed
by several collaborations, as an upper limit, a lifetime of at least
$10^{27}$~s is required for a particle with mass $10^{11}$ GeV. Thus,
the spin-$\smash{\frac{1}{2}}$ glueballs decay too
quickly for $\lambda = \mathcal{O}(1)$ since $M_{10}$ cannot be larger than
$M_4 \simeq 2 \cdot 10^{18}$~GeV. The coupling $\lambda$ can be much smaller,
though, and the spin-$\smash{\frac{1}{2}}$ glueballs may be sufficiently
stable for large enough $M_{10}$. 

Finally, we comment on throats with higher 5-form flux numbers $N_\IR$
and $N_\UV$. Since $N_\IR$ is no longer of the order 1, we have only rough 
estimates of the glueball decay rates in these cases. In the following, we
ignore this problem and assume that the glueballs are sufficiently stable. The
number $N_\UV$ is constrained by the
requirement that the Calabi-Yau has enough negative charge to compensate for
the flux. If one considers the orientifold limit of an F-theory
compactification, then this amount of negative charge is given by $\chi_4
/24$, where $\chi_4$ is the Euler number of the underlying Calabi-Yau
four-fold. Examples with $\chi_4 /24$ up to $\smash{10^4}$ are known (see
e.g.~\cite{Klemm:1996ts}) and we thus have $\smash{N_\UV \lesssim
10^4}$. Let us consider throats with maximal $N_\UV \sim 10^4$. It follows
from Eq.~\eqref{Omega} that throats at the two peaks $m_\IR \sim m_\IRc$
and $m_\IR \sim T_\RH$ of $m \cdot \eta$ can account for the dark matter if
the reheating temperature was $\sim 10^{10}$~GeV. The mass of these dark
matter candidates is $\sim  10^5$~GeV (using Eq.~\eqref{mass}) and
$\sim 10^{10}$~GeV, respectively. Together with the results from the first
part of this section (where we have chosen the other extreme with $N_\IR =
\mathcal{O}(1)$) this gives the possible range of parameters in our scenario
if the 5-form flux number is varied from its minimal to its maximal value: The
required reheating temperature is between $10^{10}$~GeV and $10^{11}$~GeV.
Moreover, the IR scale can vary between $10^5$~GeV
and $10^6$~GeV for a throat at the first peak of $m \cdot \eta$ or it is
between $10^{10}$~GeV and $10^{11}$~GeV for a throat at the second peak.

\subsection{Many throats}
\label{ManyThroats}
The distribution of vacua in the type IIB string theory landscape favours 
geometries with strongly warped regions or throats~\cite{Denef:2004ze}. For 
the class of KKLT vacua~\cite{Kachru:2003aw}, the
statistical distribution of multi-throat configurations was estimated
in~\cite{Hebecker:2006bn}. It was found that the expected number of throats
with a hierarchy $h$ larger than some $h_*$ for a given Calabi-Yau
orientifold is 
\be
\label{distribution}
\bar{n} (h > h_* | K ) = \frac{K}{3 c \log h_*},
\ee
where $K$ is the number of 3-cycles of the Calabi-Yau and $c$ is some
unknown $\mathcal{O}(1)$ constant. Using the relation $\smash{m_\IR \sim
h^{-1} N_\IR^{-1/4} M_{10}}$ and neglecting the factor $\smash{N_\IR^{-1/4}}$
for simplicity, the expected number of throats with IR scale in the range
$\smash{\hat{m}_\IR < m_\IR < \tilde{m}_\IR}$ follows from
Eq.~\eqref{distribution} and is given by
\be
\label{distribution2}
\bar{n} (\hat{m}_\IR < m_\IR < \tilde{m}_\IR |K ) =  \frac{K/3c}{\log
(M_{10}/ 
\tilde{m}_\IR)} - \frac{K/3 c}{\log (M_{10}/ \hat{m}_\IR)}.
\ee

In the following, we evaluate Eq.~\eqref{distribution2} and the resulting 
dark matter scenarios for two specific cases, always assuming that $c=1$. 
In the first case, we are optimistic about the number of 3-cycles, 
choosing $K=200$. This is a moderately high but not untypical value within 
the set of known Calabi-Yau spaces~\cite{K}. It implies a larger number 
of throats, including throats with a relatively low IR scale. In this 
case, we focus on compactifications with a moderately large volume and a 
correspondingly low 10d Planck scale, $M_{10} \sim 10^{14}$~GeV. High values 
for $M_{10}$ would lead to a long lifetime for the relatively light 
glueballs expected in this case. The observation of their decays would then 
be less likely. 

In the second case, we are conservative by choosing $K = 60$. This number 
of 3-cycles is roughly the minimal value consistent with fine-tuning of 
the cosmological constant in the KKLT construction~\cite{Hebecker:2006bn}. 
We expect only relatively short throats to be available in this case. 
Large-volume compactifications are then less interesting from a cosmological 
perspective since heavy glueballs will generically decay too quickly for low 
$M_{10}$. We thus focus on $M_{10} \sim 10^{18}$~GeV in the case $K=60$.

Obviously, many other cases, including more extreme choices of parameters,
are conceivable. However, an exhaustive study of the parameter space 
is beyond the scope of the present paper. 

If many throats are present, we can expect that the observed dark matter 
(or at least the dominant throat contribution to dark matter) will
come from throats with $m_\IR$ near one of the two maxima of 
$m \cdot \eta$ (cf.~Fig.~\ref{fig:plot}). As before, we will simplify 
the analysis by using $N_\UV \sim 10$ for long throats ($m_\IR\sim 10^6$ 
GeV) and $N_\UV\sim 4$ for short throats ($m_\IR\sim 10^{11}$~GeV). If 
dark matter is indeed due to such throats, the reheating temperature 
has to be $\smash{T_\RH \sim 10^{11}}$~GeV. Note that, in general, the 
situation might be more complicated: For example, a throat with an IR
scale far away from the maxima but with very large $N_\UV$ (recall that 
$N_\UV$ is fairly arbitrary if we do not insist on $N_\IR\sim 1$) may provide 
the dominant contribution to dark matter (cf.~Eqs.~\eqref{Tpt} 
and~\eqref{Omega}).

In the first example with $M_{10} \sim 10^{14}$~GeV, the lifetime of glueballs
from a throat with $m_\IR \sim 10^{11}$~GeV is much too short to be a good
dark matter candidate (see below). Therefore, we focus on the maximum of $m
\cdot \eta$ near $m_\IR\sim \smash{10^6}$~GeV. The expected number of throats
with this IR scale is
\be
\bar{n} \left(5 \cdot 10^5 \text{ GeV} <  m_\IR < 5 \cdot 10^6 \text{ GeV}
\right) \,
\simeq \, 0.5.
\ee
Thus, a significant fraction of the vacua has a throat which yields the
right amount of dark matter for $\smash{T_\RH \sim 10^{11}}$
GeV.\footnote{
In deriving the energy transfer rates in Eqs.~\eqref{tr1} and \eqref{etr}, we
have assumed that the reheating temperature is smaller than the
compactification scale, i.e.~$\smash{T_\RH < L^{-1}}$. Using the relation $M_4
\sim L^3 M_{10}^4$, the compactification scale is $\smash{L^{-1} \sim
10^{12}}$~GeV for a 10d Planck scale $M_{10} \sim 10^{14}$~GeV. Thus, the
assumption is still fulfilled for such a low 10d Planck scale and a reheating
temperature $T_\RH \sim 10^{11}$~GeV.} 
Certain partial lifetimes of the glueballs have been discussed in
Sect.~\ref{SingleThroat}. In addition, we now expect
\be
\bar{n} \left( m_\IR \lesssim 10^5 \text{ GeV} \right) \, \simeq \, 3.5
\ee
throats with IR scales smaller than $\smash{10^6}$~GeV which provide another 
decay channel for the glueballs. These throats can have a large
number $G$ of degrees of freedom. Therefore, we have to check whether the
lifetime of the dark matter glueballs is still larger than the current age of
the
universe. If we denote the 5-form flux number at the UV end of the $i$th
throat by $N_i$, we have
\be
\label{dof}
G \, = \,  \sum_i N^2_i,
\ee
where the sum runs over all throats with IR scales smaller than 
$\smash{10^6}$~GeV. Using Eq.~\eqref{scalardr} with $N^2 = G$, the partial
lifetime of the glueballs for decays to these throats is
\be
G^{-1} \, 10^{32} \text{ s}.
\ee
According to the discussion at the end of Sect.~\ref{SingleThroat}, we
expect $G$ to be somewhere in range of $\smash{10^2}$ to $\smash{10^8}$.
Thus, even for maximal $G$ this partial lifetime is much larger than the
current age of the universe and the glueballs are still a good dark matter
candidate. 

There are also throats with IR scales larger than $10^6$~GeV. Since throats
with IR scales larger than the reheating temperature are not heated for
kinematic reasons, we are interested in throats between $10^7$~GeV and
$10^{11}$~GeV. Their average number is
\be
\bar{n} \left( 10^7 \text{ GeV} \lesssim m_\IR \lesssim  
10^{11} \text{ GeV} 
\right) \, \simeq \, 8.6.
\ee
Glueballs from these throats have shorter lifetimes than the dark matter
glueballs. The abundance of particles which decay to the standard model
with a lifetime between $\smash{10^{-2}}$~s and  $\smash{10^{12}}$~s
is severely constrained by nucleosynthesis~\cite{UP2,UP1,Kawasaki:2004qu}.
Therefore, we have to check whether the decaying glueballs fulfill
these bounds.\footnote{For lifetimes larger than $10^{12}$ s, bounds from the
diffuse $\gamma$-radiation are important~\cite{KribsRothstein}. These
are not relevant in this example.}

We restrict ourselves to decays of scalar glueballs. The discussion
can then be easily extended to include the fermionic glueballs. Since
the fermionic glueballs decay to the standard model via the
operator in Eq.~\eqref{vertex3}, nucleosynthesis may give a bound on the
coupling strength $\lambda$. Scalar glueballs have three important decay
channels: They decay to two gravitons, to throats with lower IR scales and to
the standard model. The total decay rate is the sum of the three
corresponding decay rates. For a throat at $10^{11}$~GeV, the total decay rate
is dominated by decays to throats with lower IR scales. Denoting by
$G$ the combined number of degrees of freedom of throats with $m_\IR <
10^{11}$~GeV and using Eq.~\eqref{drSpin2}\footnote{
In deriving
the decay rate in Eq.~\eqref{scalardr}, we have assumed that the mass $m_\tau$
of the modulus which mediates the decay is larger than the mass $m_\IR$ of the
decaying glueball. Using Eq.~\eqref{mtau}, we have $m_\tau \sim 10^{10}$~GeV 
for $M_{10} \sim 10^{14}$~GeV. For a throat with $m_\IR \sim 10^{11}$~GeV, we
therefore have to use the unsuppressed decay rate in Eq.~\eqref{drSpin2}.},
the glueballs from such a throat decay already after $G^{-1} 10^{-7}$~s. Since
this lifetime is shorter than $10^{-2}$ s, these glueballs do not affect
nucleosynthesis. Similarly, glueballs from a throat at $10^{10}$~GeV do not
live long enough to be relevant for nucleosynthesis. 

The lifetime becomes larger for throats with lower IR scales. We consider a
throat at $10^7$~GeV. The total decay rate is then dominated by decays to two
gravitons since $G \lesssim 10^8$.
Using Eq.~\eqref{scalardr}, the corresponding lifetime is
approximately $10^{10}$~s. Let 
us denote the mass density over entropy density of the fraction of glueballs
that have decayed to the standard model sector by $ m\cdot \eta_\dec$.
Successful nucleosynthesis requires
that~\cite{Kawasaki:2004qu}
\be
\label{bound1}
m \cdot \eta_\dec \lesssim 10^{-14}  \text{ GeV}
\ee
for particles that decay $10^{10}$~s after reheating.\footnote{
The bounds on $m
\cdot \eta_\dec$ were derived for particles with masses in the range of $10^
2$~GeV to $10^4$~GeV. At $10^{10}$ s, the bound is approximately independent
of the particle mass. We therefore believe that it is a reasonable
approximation to extrapolate this bound to particles of mass $10^7$ GeV.} To
estimate $m
\cdot\eta_\dec$, recall that $m \cdot \eta$ has maxima at IR
scales $m_\IRc \sim
10^6$~GeV and $T_\RH \sim 10^{11}$~GeV (see Fig.~\ref{fig:plot}). Calculating
$m \cdot \eta$ at these IR scales from Eq.~\eqref{energydensity2}, we have 
\be
\label{upperbound}
m \cdot \eta \lesssim 10^{-9} \text{ GeV}.
\ee 
Using this upper value
and taking the branching ratio for decays to the standard model sector into
account, we get $m \cdot \eta_\dec \lesssim 10^{-19}$~GeV for the
fraction of scalar glueballs that may have affected nucleosynthesis.
Thus, the bound in Eq.~\eqref{bound1} is clearly fulfilled.
Similarly, one can check that scalar glueballs from throats at $10^8$~GeV
and $10^9$~GeV fulfill the bounds from nucleosynthesis. We conclude that
glueball decays from
throats between $10^7$~GeV and $10^{11}$~GeV do not destroy nucleosynthesis. 

In the second example, it is unlikely to find throats with 
$m_\IR\sim 10^6$~GeV. Thus, we focus on the maximum of $m \cdot \eta$ near 
$m_\IR\sim \smash{10^{11}}$~GeV. According to Eq.~\eqref{distribution2}, the 
expected number of throats with IR scale between $\smash{10^{10}}$~GeV and 
$\smash{10^{11}}$~GeV is
\be
\bar{n} \left( 10^{10} \text{ GeV} \lesssim m_\IR \lesssim 10^{11} \text{
GeV} 
\right) \, \simeq \, 0.3\,.
\ee
Thus, a significant fraction of the vacua has a throat in this range of IR
scales. A reheating temperature of the order of $\smash{ 10^{11}}$~GeV would
give the right amount of dark matter not only for a throat at
$\smash{10^{11}}$~GeV
but also for a throat at $\smash{10^{10}}$~GeV: As one can see from
Fig.~\ref{fig:plot} (the thin curve), we expect the function $m \cdot \eta$ to
decrease rather slowly with the IR scale near the peak at $m_\IR\sim T_\RH$.
In the worst case, $m \cdot \eta$ is proportional to $m_\IR$. From the 
discussion in Sect.~\ref{te}, we also know that $m \cdot \eta$ is
proportional to $T_\RH^3$. Thus, in order to compensate the decrease in $m 
\cdot \eta$ if $m_\IR$ is lowered by one order of magnitude, $T_\RH$ only 
has to be raised by an $\mathcal{O}(1)$ factor.

Certain partial lifetimes of the glueballs from such a throat have been
discussed in Sect.~\ref{SingleThroat}. In addition, there may be throats with
lower IR scales which provide new decay channels for the glueballs. Their
expected number is 
\be
\bar{n}\left(m_\IR \lesssim 10^9 \text{ GeV} \right) \, \simeq \, 1.1.
\ee
We denote the number of degrees of freedom of this sector by $G$. As before,
we have to check that the dark matter glueballs do not decay too quickly to
this sector. Using Eq.~\eqref{scalardr}, the partial lifetime for this decay
channel is
\be
\label{lt2}
G^{-1} \, 10^{29} \text{ s} \quad \text{to} \quad G^{-1} \, 10^{20} \text{ s}
\ee
for glueball masses in the range of $\smash{10^{10}}$~GeV to $\smash{10^{11}}$
GeV. For large $G$, this lifetime is larger than the current age of
the universe only for a throat around $\smash{10^{10}}$~GeV. If $G$ is small
or if there is no throat with lower IR scale, also a throat around
$10^{11}$~GeV would give a sufficiently stable dark matter candidate.

\section{Further issues related to supersymmetry breaking}
\label{extra}
Up to now, we have neglected decays mediated by the K\"ahler moduli. This is 
justified if the K\"ahler moduli are sequestered from the throat sectors. 
The latter assumption follows from the interpretation of a Calabi-Yau 
orientifold with a long throat as supersymmetric Randall-Sundrum model in 
which the K\"ahler moduli are localized on the UV brane~\cite{Brummer:2005sh}. 
The sequestering assumption in this 5d framework~\cite{Randall:1998uk} has 
been widely accepted and has also been used in the context of type-IIB models 
with strongly warped regions (see e.g.~\cite{Choi:2006bh,Brummer:2006dg} as 
well as the detailed discussion of~\cite{Kachru:2007xp} and refs.~therein). 

Let us restrict ourselves to the universal K\"ahler modulus. 
We denote the corresponding chiral superfield by $T$ and a chiral glueball 
superfield from the throat sector by $X$. The Lagrangian can be written in 
standard ${\cal N}=1$ supergravity form
\be
\mathcal{L} \, = \, \int d^4 \theta \; \varphi \bar{\varphi} \; \Omega + 
\left(\int d^2 \theta \; \varphi^3 \; W + \text{ h.c.} \right)\,,
\ee
where $\varphi = 1 +\theta^2 F_\varphi$ is the chiral compensator, $\Omega$ 
is the kinetic function and $W$ is the superpotential. The sequestering 
assumption~\cite{Randall:1998uk} states that
\begin{equation}
\begin{split}
\label{sequestered}
\Omega(X, \bar{X}, T, \bar{T}) \; & = \; \Omega(X, \bar{X}) \, + \, 
\Omega(T, \bar{T}) \\
W (X, T) \; & = \; W(X) \, + \, W(T).
\end{split}
\end{equation}
In particular, terms mixing the superfields $T$ and $X$ appear neither in the 
kinetic part nor in the superpotential. Thus, since the universal K\"ahler 
modulus does not mix with the glueballs, it cannot mediate their decays to 
other sectors. Even if the sequestered form of Eq.~\eqref{sequestered} turns 
out to be violated, we expect that the cross-couplings are much more 
suppressed than the mixing vertex of Eq.~\eqref{vertex} between glueball and 
dilaton. The effect of the K\"ahler moduli in mediating glueball decays is 
then still negligible.

We have also not yet considered decays mediated by the gravitino. It may turn 
out that the fermionic glueball mixes with the gravitino which thus mediates 
its decay to other sectors. An additional process is the decay of the heavier 
superpartner to the lightest glueball by the emission of a gravitino. This 
follows from the process shown in Fig.~\ref{fig:decay1} by replacing the 
virtual spin-2 glueball by a virtual spin-$\smash{\frac{3}{2}}$ glueball, 
the outgoing graviton by a gravitino and one of the bosonic glueballs by a 
fermionic glueball. If the emitted gravitino is heavier than the decaying 
glueball, the gravitino is off-shell and must in turn decay to the standard 
model or another throat. It is not immediately clear, how strongly the 
propagator of the gravitino suppresses the corresponding decay rate, i.e. 
with which power the gravitino mass enters. In addition, the gravitino can 
be considerably lighter than the dilaton and complex-structure moduli. It may 
therefore turn out that we get a strong bound on the gravitino mass in our 
scenario. This would make it more probable that the decaying glueballs lead to 
a detectable signal. This is an interesting topic for future investigations. 

Finally, let us consider the limit of light gravitinos or, equivalently, 
supersymmetry broken at a low scale. If the gravitino is lighter than the 
glueballs, the spin-$\smash{\frac{1}{2}}$ glueballs may decay to a 
graviton and a gravitino as discussed at the end of Sect.~\ref{processes}.
The scalar glueballs decay to two gravitons with the same rate. We have seen 
in Sect.~\ref{SingleThroat} that the scalar glueballs with mass 
$\smash{10^6}$~GeV have a lifetime shorter than the current age of the 
universe. This lifetime would now also apply to the fermionic glueballs.
But the partial lifetime for this decay channel is proportional to
$m_\IR^{-5}$ according to Eq.~\eqref{dr2}. Thus, this problem does not occur if
the infrared scale is somewhat lower than our `optimal' value of $10^6$~GeV.
In this case, throat dark matter is still possible even if the SUSY breaking
scale is low. 

Many of the decay processes discussed in this and previous sections depend on 
the spectrum of the lightest glueballs from the throat sector. This spectrum
depends crucially on the pattern of supersymmetry breaking in the throat, to
which we now turn. Motivated by sequestering, we assume that supersymmetry
breaking is communicated to the lightest glueball multiplet $X$ only by the
F-term $F_\varphi$ of the chiral compensator.\footnote{
Actually,
the situation might be more complicated since the lightest glueball multiplet 
couples strongly to heavier glueballs, which are themselves affected by 
supersymmetry breaking and which might therefore influence the mass-splitting 
of the lightest multiplet in a non-negligible way. We intend to return to 
this point in future investigations and view the present calculation  
only as a reasonable first guess.
}
The relevant part of the effective Lagrangian Eq.~\eqref{sequestered} is 
\be
\label{susybreaking}
\mathcal{L} \, \supset \, \int d^4 \theta \; \varphi \bar{\varphi} \, X 
\bar{X} + \left( \int d^2 \theta \; m X^2 \varphi^3 + \text{ h.c.} \right).
\ee
If supersymmetry is broken, the $F$-term of the chiral compensator gets a vev 
$\smash{\langle F_\varphi \rangle}$. Since we can expect this vev
to be of the same order of magnitude as the gravitino mass, the limit of a 
light gravitino corresponds to $\smash{\langle F_\varphi \rangle \ll m}$. 
We can expand $X$ in components and split the lowest component of $X$ into 
real and imaginary parts. Inserting this expression in 
Eq.~\eqref{susybreaking} and diagonalizing the resulting mass matrix for 
the real and imaginary part, one finds two scalar eigenstates with masses
\be
m_{1,2}^2 \, = \, 4 \, m^2 \pm \, 2 \, m \, |\langle F_\varphi \rangle|.
\label{split}
\ee
Moreover, the mass of the fermion is $2 m$ and receives no contribution from 
$\smash{\langle F_\varphi \rangle}$. Therefore, one scalar glueball is lighter 
than its former spin-$\smash{\frac{1}{2}}$ superpartner and the mass splitting
is $|\langle F_\varphi \rangle|/2$. Depending on the precise relation between 
$\langle F_\varphi \rangle$ and the gravitino mass, the 
spin-$\smash{\frac{1}{2}}$ glueball may or may not decay to the lighter scalar
glueball by emission of a gravitino. If this decay is kinematically not 
allowed, the spin-$\smash{\frac{1}{2}}$ glueball may still decay by mediation 
of a gravitino to standard model particles and the lighter scalar glueball.
The gravitino propagator gives no suppression of the decay rate in this 
case and the decay rate is given by Eq.~\eqref{drSpin2}. For a throat with
$\smash{m_\IR \sim 10^6}$~GeV and $N_\IR\sim {\cal O}(1)$, the partial
lifetime of the spin-$\smash{\frac{1}{2}}$ glueballs for this decay channel is
\be
\tau \, \sim \, 10^{15} \text{ s}.
\ee
Again, this is shorter than the current age of the universe. Since decays of
this kind are constrained by diffuse $\gamma$-ray measurements, a partial
lifetime larger than $\sim 10^{26}$~s is actually required. Since the partial
lifetime is again proportional to $m_\IR^{-5}$ (cf.~Eq.~\eqref{drSpin2}),
also this problem can be avoided with a lower IR scale. As we have explained
in Sect.~\ref{SingleThroat}, this requires a higher reheating temperature. 

Let us finally note that Eq.~(\ref{split}) is obviously not applicable if
$\smash{\langle F_\varphi \rangle \gg m}$. In this case, we can analyse the 
situation from the perspective of a chiral superfield with vanishing mass, 
i.e. we consider the limit $m\to 0$. The theory then possesses a chiral 
symmetry which ensures the masslessness of the fermion even in the presence 
of SUSY breaking. Thus, in analogy to the matter superfields of the minimal 
supersymmetric standard model, we expect that the scalar glueballs will
be heavier than the fermions if supersymmetry breaking in the throat is 
a large effect relative to the supersymmetric mass term.

\section{Summary}
\label{summary}
Strongly warped regions or throats are a common feature of the type 
IIB string theory landscape. KK modes whose wavefunctions are localized in a 
throat have redshifted masses, allowing for their production after reheating 
in the standard model sector, even if the reheating temperature is not very 
high. In addition, these KK modes are only very weakly coupled to the rest 
of the compact manifold, potentially resulting in a very long lifetime. These 
properties make the KK modes an interesting dark matter candidate. 

We have considered a setup in which the standard model lives in the unwarped 
part of a compact manifold, which in addition has a certain number of 
throats. To be conservative, we have assumed that reheating only takes place 
in the standard model sector. Even under this minimal assumption, the 
throats are heated up by energy transfer from the hot standard model plasma. 
In Sect.~\ref{et}, we have determined the energy density deposited
in a throat, using our result for the corresponding energy transfer rate from
a previous paper~\cite{hotthroats}.

Throats have a dual description as a strongly coupled gauge theory with a 
large number of colours. From this gauge theory point of view, the energy 
density in a throat is initially in the form of gauge theory states 
with energy of the order of the reheating temperature. These states 
subsequently settle into a certain number of lighter glueballs with some
distribution 
of kinetic energies. The knowledge of this distribution is important since 
it determines whether the glueballs thermalize and whether (and for how 
long) the energy density scales like radiation or like matter. Since we are 
at present unable to determine this distribution of kinetic energies, we 
have considered two extremal cases in Sect.~\ref{te}. In the first case, 
the initial gauge theory state settles into a large number of glueballs 
with kinetic energies of the order of their mass. In the second case, the 
decay products of the initial state are an $\mathcal{O}(1)$ number of 
glueballs which accordingly have kinetic energies of the order of the 
reheating temperature. 

It turns out that the gauge theory thermalizes in both extremal cases if the
energy density in this sector is above the critical energy density for
deconfinement. The energy density thus scales like radiation with the
expansion of the universe until the confinement phase transition takes place.
Afterwards, it scales like matter. Taking this scaling into account, we have
determined the contribution of the throat sector to the total energy density
of the universe at our epoch. We have also determined this contribution for
the case that the initial energy density in the throat sector is below the 
critical energy density and found that it differs considerably 
between the two extremal cases. We expect the true behaviour in this region 
of parameter space to be in between the two extremal cases. 

After the confinement phase transition, different types of glueballs with 
mass of the order of the confinement scale $m_\IR$ are formed. Similarly, 
if the gauge theory does not thermalize, the initial gauge theory states
created at reheating settle into a certain number of light glueballs. As we
have shown in Sect.~\ref{processes}, these glueballs quickly decay to a
lightest state and its superpartner. On the basis of a number of papers
devoted to the spectrum of the KS gauge
theory~\cite{Krasnitz:2000ir,Caceres:2000qe,
Amador:2004pz,Firouzjahi:2005qs,Noguchi:2005ws,Dymarsky:2006hn,Berg:2006xy,
Dymarsky:2007zs,Benna:2007mb}, we expect these lightest states to be a scalar
glueball 
and its spin-$\smash{\frac{1}{2}}$ superpartner. The glueballs couple (very 
weakly) to the standard model and other throats and can thus decay to these 
sectors. In Sect.~\ref{DecaytoSM}, we have applied our results for this 
decay~\cite{hotthroats} to scalar glueballs. A crucial difference with 
respect to our previous analysis is the fact that the bulk fields which 
mediate the decay are massive in a flux background. Depending on the mass of 
the decaying glueballs, this can lead to a significant suppression of the 
corresponding decay rate. In addition, scalar glueballs can decay to two 
gravitons. We have determined the corresponding rate in Sect.~\ref{processes}.

Similarly, spin-$\smash{\frac{1}{2}}$ glueballs can in principle decay to 
a graviton and a gravitino. To get a stable dark matter candidate, we are
mainly interested in setups where such decays are kinematically forbidden
due to a heavy gravitino. This requires that the supersymmetry breaking scale
is larger than the mass of the glueball. It may also mean that the
superpartners of standard model particles are heavier than the
glueballs. If R-parity is conserved, most decay channels
of spin-$\smash{\frac{1}{2}}$ glueballs to the standard model sector involve
such a superpartner and are therefore kinematically forbidden. In
Sect.~\ref{DecaytoSM}, we have identified an operator which does not involve a
superpartner and which would allow the decay of spin-$\smash{\frac{1}{2}}$
glueballs to a Higgs and a lepton. If present, this operator would give the
dominant decay channel even for maximally broken R-parity. Using the
corresponding vertex, we have determined the decay rate of
spin-$\smash{\frac{1}{2}}$ glueballs to the standard model.

In Fig.~\ref{fig:plot}, we have plotted the contribution of a throat sector 
to the total energy density of the universe for fixed reheating temperature 
$T_\RH$ and as a function of the IR scale $m_\IR$. We expect this function to 
have two maxima at IR scales $ m_\IRc$ and $T_\RH$. For a throat with IR 
scale $m_\IRc$, the dual gauge theory thermalizes precisely to the phase 
transition temperature and therefore the energy density scales like matter 
immediately after reheating. KK modes in throats with IR scale $T_\RH$, on 
the other hand, are so massive that they become nonrelativistic immediately 
after reheating and the energy density again scales like matter afterwards. 

One of the main underlying ideas of our analysis is the generic 
presence of throats in the type IIB landscape. It is then probable to have 
throats with these optimal IR scales in a given compact manifold. For 
simplicity, we have first discussed a scenario with a single throat in 
Sect.~\ref{SingleThroat}. We have found that, in many cases, KK modes in 
throats with IR scale $T_\RH$ have a decay rate which is too high for them 
to be a good dark matter candidate. However, if the gravitino is very heavy 
(high-scale SUSY breaking) and certain operators connecting the 
standard model and the moduli sector are suppressed, the fermionic glueballs 
may nevertheless survive and play the role of dark matter. 

The more promising case is that of throats with IR scale $m_\IRc$, to which we
now turn. In this case, it follows immediately from our results of
Sect.~\ref{te} that a reheating temperature of $10^{10}$~GeV to $10^{11}$~GeV
leads to the
right amount of glueballs to account for the observed dark matter. The
critical IR scale $m_\IRc$ is a function of $T_\RH$. After having fixed 
$T_\RH$, we find a mass for our dark matter candidate which is between
$\smash{10^5}$~GeV and $10^6$~GeV. In Sect.~\ref{ManyThroats}, we have used
results from Ref.~\cite{Hebecker:2006bn} on the distribution of throats 
in the landscape to consider a scenario with a large number of throats. We 
have found that there are setups in which the probability of having a throat 
with the required IR scale is large. The only free 
parameter which has then to be fixed is the reheating temperature. 

Our dark matter scenario may lead to some interesting observable signatures. 
The dark matter glueballs decay to the standard model with a very low,
but non-negligible rate. The decays produce photons to which experiments like 
GLAST or HESS may be sensitive. It turns out that the decay rates depend on
two parameters: The 10d Planck mass $M_{10}$ enters via the flux
stabilized mass of fields which mediate the decay. Moreover, the
decay rate of spin-$\smash{\frac{1}{2}}$ glueballs depends on the coupling
strength $\lambda$ of the aforementioned operator which allows their decay to
a lepton and a Higgs. In Sect.~\ref{SingleThroat}, we have identified two
interesting scenarios: 

If $\lambda$ is of the order 1, $M_{10}$ has to be very large in order to get
a sufficiently stable dark matter candidate. Namely, for a lower 10d Planck
scale, the spin-$\smash{\frac{1}{2}}$ glueballs decay to the standard model
with a rate which is in conflict with measurements of the diffuse
$\gamma$-radiation. On the other hand, the 10d Planck scale cannot be larger
than the 4d Planck scale. This makes it more probable that the lifetime of
spin-$\smash{\frac{1}{2}}$ glueballs is in a range that can be probed with
new, more sensitive experiments like GLAST. If the scenario with $\lambda
=\mathcal{O}(1)$ is realized in nature, one can hope to detect a signal from
the decaying glueballs in the near future.

If $\lambda$ is much smaller than 1, a lower $M_{10}$ still leads to
sufficiently stable spin-$\smash{\frac{1}{2}}$ glueballs. For a
low 10d Planck scale, also the decay of scalar glueballs can become relevant
for detection.
In contrast to fermionic glueballs, scalar glueballs can decay directly to
two photons. This decay channel leads to a sharp $\gamma$-line at an energy of
$10^5$~GeV to $10^6$~GeV which could be detected with experiments like HESS.
In addition, the scalar glueball decays also produce a continuous
$\gamma$-ray spectrum to which e.g.~GLAST may again be sensitive. If the 
scalar glueballs make up an $\mathcal{O}(1)$ fraction of the dark matter at
our epoch, a 10d Planck scale of the order of $10^{13}$~GeV
would allow for a detection of both signals in the near future. Such a
10d Planck scale corresponds to a compactification radius
of the order of just $ 50\,l_{\bf string}$, which is not extremely large. 

Another interesting effect is the decay of the scalar glueballs into two 
gravitons which may happen before our epoch. Since we consider 
a setup in which the corresponding decay of the spin-$\smash{\frac{1}{2}}$ 
glueballs into a graviton and a gravitino is kinematically forbidden, we still
have a sufficiently stable dark matter candidate. For reasons that we have
explained in Sect.~\ref{processes}, we expect the scalar glueballs
to have a lower abundance than the spin-$\smash{\frac{1}{2}}$ glueballs.
The total dark matter abundance would then only change by a small factor 
when the scalar glueballs decay. It is possible, however, that the scalar
glueballs have a comparable or even higher abundance than the
spin-$\smash{\frac{1}{2}}$ glueballs. The change in the dark matter abundance
could then be enormous. It would be interesting to consider the implications of
this scenario for cosmology.

In a setup with a large number of throats one also expects throats with IR
scales larger than $10^5$~GeV to $10^6$~GeV (the IR scale of the throat
which provides the dark matter). The glueballs from these throats have
shorter lifetimes and may decay already during nucleosynthesis. Therefore,
successful nucleosynthesis may impose further constraints on our scenario. For
a particular example, we have checked in Sect.~\ref{ManyThroats} that the
glueball decays do not affect nucleosynthesis due to a low branching ratio to
the standard model. Examples are conceivable, however, in which
bounds from nucleosynthesis are only marginally fulfilled. It would then be
interesting to look for traces of glueball decays in the abundances of light
elements. Unfortunately, we are at present unable to determine the
contribution of throats with the aforementioned range of IR scales to the
total energy density of the universe with sufficient precision. Progress in
this direction requires a detailed understanding of hadronization in a
strongly coupled gauge theory, which we therefore view as a further
interesting topic for future research.

\vspace*{.4cm}
\noindent
{\bf Acknowledgements}:\hspace*{.5cm}We would like to thank F.~Br\"ummer,
W.~Buchm\"uller, T.~Dent, J.~Ellis, W.~Hofmann, K.~Sigurdson and M.~Trapletti
for
helpful discussions.

\end{document}